\documentclass[12pt]{iopart}
\usepackage{graphicx}
\usepackage{bm}
\usepackage{color}
\usepackage{url}
\usepackage{bbold}
\usepackage{amssymb}
\usepackage{amsfonts}
\usepackage{tensor}

\begin{document}
\def\ee{{\mathrm e}}
\newcommand{\ac}{\cal S}
\newcommand{\acEH}{{\cal S}_{\rm EH}}
\newcommand{\lag}{{\mathscr L}}
\newcommand{\emT}{\mathpzc{T}}
\newcommand{\aeT}{\emT^{\ae}}
\newcommand{\ADM}{{\text{\sc adm}}}
\newcommand{\khor}{\text{\sc kh}}
\newcommand{\uhor}{\text{\sc uh}}
\newcommand{\Gae}{G_{\ae}}
\newcommand{\GN}{G_\text{\sc n}}
\newcommand{\met}{\mathsf{g}}
\newcommand{\hatmet}{\hat{\mathsf{g}}}
\newcommand{\Rie}{\mathpzc{R}\,}
\newcommand{\Ric}{\mathpzc{R}\,}
\newcommand{\EinG}{\mathpzc{G}}
\newcommand{\Sph}{\boldsymbol{\mathcal{B}}}
\newcommand{\tn}{\tensor}
\newcommand{\dl}{\partial}
\newcommand{\Dl}{\nabla}
\newcommand{\LieD}{{\large\text{\pounds}}}
\newcommand{\Mpl}{M_{\rm Pl}}

\title[Tests of Lorentz invariance]{Tests of Lorentz invariance: a 2013 update}

\author{S. Liberati}

\address{SISSA, Via Bonomea 265, 34136 Trieste, Italy\\
INFN, Via Valerio 2, 34127 Trieste, Italy.}
\ead{liberati@sissa.it}

\begin{abstract}
We present an updated review of Lorentz invariance tests in Effective field theories (EFT) in the matter as well as in the gravity sector. After a general discussion of the role of Lorentz invariance and a derivation of its transformations along the so called von Ignatovski theorem, we present the dynamical frameworks developed within local EFT and the available constraints on the parameters governing the Lorentz breaking effects. In the end, we discuss two specific examples, the OPERA ``affaire" and the case of Ho\v{r}ava-Lifshitz gravity. The first case will serve as an example, and a caveat, of the practical application of the general techniques developed for constraining Lorentz invariance violation (LIV) to a direct observation potentially showing these effects. The second case will show how  the application of the same techniques to a specific quantum gravity scenario has far fetching implications not foreseeable in a purely phenomenological  EFT approach.

\end{abstract}
\submitto{\CQG}
\maketitle

\section{Introduction}

Quantum gravity has been a frustrating endeavour for more than sixty years and in spite of many technical progresses it is a program far from completion.
We have nowadays many approaches to quantum gravity, string theory, loop quantum gravity, causal dynamical triangulations, causal sets, quantum graphity, just to name a few, but these are radically different approaches, driven by underlying different guiding ideas and even general lessons about the nature of spacetime beyond the Planck scale are difficult to draw. Why so? The answers probably lies in the fact that Quantum Gravity has been a collective effort in theoretical physics like no other before, a search without lighthouses i.e. without observations to guide us. Indeed, we expect QG effects at experimentally/observationally accessible energies to be extremely small, due to suppression by the Planck scale $M_{\rm Pl} \equiv \sqrt{\hbar c/G_{\rm N}}\simeq 1.22\times 10^{19}~\mbox{GeV}/c^{2}$. In this sense it has been considered (and it is still considered by many) that only ultra-high-precision (or Planck scale energy) experiments would be able to test quantum gravity models.

Nonetheless, no serious theoretical physicist or cosmologist can nowadays be satisfied of our current theoretical toolbox. Indeed, the standard model of particle interactions (SM) and classical General Relativity (GR) are extremely powerful tools and are rightly listed among the most impressing achievements of mankind. Yet, these theories are not enough for answering many pressing questions. Our observations are starting to probe physics just $10^{-35}$ seconds after the Big Bang asking for theories able to set the initial conditions of our Universe. Furthermore, our cosmology makes sense within our current theoretical framework only at the price of a very preposterous choice of matter-energy content, we do understand only about 4\% it. Even worse GR fails to be a predictive theory in many interesting and relevant situations. Indeed, many solutions of Einstein's equations are singular in some region, not least the very beginning of our Universe or the interior of black holes (for which we do have nowadays a solid evidence of existence). Moreover, there are honest classical solutions of the Einstein equations that contain closed time-like curves, which would allow traveling back and forth in time with the associated causal paradoxes.  Finally, the problem of black-hole evaporation considered just within the framework of semi-classical gravity clashes with quantum mechanical unitary evolution.  

The conclusion is one and the same: we need nowadays much more than in the past a theory of quantum gravity, something able to describe the nature and the dynamics of spacetime beyond the Planck scale. So, even more than in the past any attempt to put boundaries and constraints on our intuitions/ideas/speculations about what lies behind the Planck scale are necessary and should not be dismissed as simple minded or premature.

The broad and multi-faceted field of ``quantum gravity phenomenology" has represented in the last twenty years a difficult, controversial and painstaking attempt to do this. To grasp whatever could be grasped by observations and to tests any (more or less crazy) idea could be advanced and tested about Planck scale effects and their low energy imprint on reality. 

In fact, quantum gravitational models beyond GR  have shown that there can be several of what we term low energy ``relic signatures'' of quantum gravitational effects which would lead to deviation from the standard theory predictions (SM plus GR) in specific regimes. Some of these new phenomena, which comprise what is often termed ``QG phenomenology'', include tests of
\begin{itemize}
\item Quantum decoherence and state collapse \cite{Mavromatos:2004sz}
\item QG imprint on initial cosmological perturbations \cite{Weinberg:2005vy}
\item Cosmological variation of couplings \cite{Damour:1994zq,Barrow:1997qh}
\item TeV Black Holes, related to extra-dimensions \cite{Bleicher:2001kh}
\item Planck scale spacetime fuzziness \cite{AmelinoCamelia:1998ax}
\item Generalized uncertainty principle \cite{Garay:1994en,Hossenfelder:2012jw,marin2012gravitational}
\item Violation of discrete symmetries \cite{Kostelecky:2003fs}
\item Violation of space-time symmetries \cite{Mattingly:2005re}
\end{itemize}
This review will focus on a special subset of the last item i.e. Lorentz violations. 

\subsection{Why testing Lorentz invariance?}

Lorentz invariance lies at the roots of both the SM and GR and as we shall see soon it is really an outcome of our deepest understanding of the nature of space and time. So why we should sacrifice it on the altar of quantum gravity, isn't this a too high price to pay? The answer to this question is twofold. First, there is a practical issue: exactly because Lorentz Invariance (LI) is a cornerstone of our current understanding of reality we do have a compelling duty to test it as far as we can. Secondly there are good reasons to investigate about the compatibility of this cherished symmetry with Planck scale physics. In fact, many models of quantum gravity involve some form of discretisation of spacetime, something that a priori is hard to reconcile with e.g. boost invariance. Of course, one can envisage a discretisation of spacetime which is by construction LI, like e.g.~in causal sets.  However Lorentz invariance of the sub-Planckian structure of a QG model does not per se guarantee the preservation Lorentz invariance at intermediate scales. Let us dwell a bit more carefully on this point. 

All our models of QG have to solve a crucial issue related to the emergence of a classical spacetime at sufficiently low scales. It is often assumed that as far as the underlying structure does not involved a preferred frame so the emergent structures will. This is however not supported by simple toy models that one can construct in this sense. To cite just a simple example, it was shown within the so called Analogue Gravity framework \cite{Fagnocchi:2010sn, AG:2011} that a system of relativistic bosons that undergoes a Bose-Einstein condensation is characterised by low energy collective excitations (quasi-particles) that experience the condensate as an effective spacetime and are characterised by a relativistic dispersion relation with the speed of light replaced by the speed of sound in the BEC, $\omega^2=c^2_s k^2$. However, the same excitations at high energy have cumbersome dispersion relations that break what we might call this ``acoustic" LI and interpolate finally with a UV fully relativistic dispersion relation this time characterised by the speed of light as the limit speed. While this toy model cannot be accounted in any way as a model of quantum gravity (as it shows an emergent spacetime but cannot reproduce and emergent GR dynamics~\footnote{See however \cite{Girelli:2008gc,Girelli:2008qp,Sindoni:2010eu} for analogue gravity or an analogue gravity inspired models showing possible emergence of gravitational-like dynamics in these systems.}) it is definitely making a point about the fact that an underlying Lorentz invariance of the system is not sufficient (and could be not even necessary) in order to provide an almost exact Lorentz invariant physics in the emergent spacetime.

\section{The von Ingnatovski theorem}
\label{vIgnth}

In discussing the viable test of Lorentz invariance it might be fruitful to take some time to express first what we are actually probing when we cast the aforementioned constraints. It is in fact often associated with LI the observation that the speed of massless particles (i.e. the speed of light) is the same independently of the chosen reference system in which the measurement is performed.

This is true and false at the same time. While it is surely true that a route to the derivation of Lorentz transformations is to assume a priori the invariance of the speed of light, it was also soon understood \cite{ignatowsky,ignatowsky1,ignatowsky2,ignatowsky3,ignatowsky4} that no such assumption was needed and that indeed the basic transformations of Special Relativity could be derived starting from basic assumptions about the fundamental nature of spacetime.

In this sense a first step consists in realising that no metaphysical, abstract, meaning needs to be associated to spacetime. Indeed, this can be defined operationally by replacing the abstract notions of ``time" and ``space" with the physical notions of ``duration" (of a physical phenomenon as measured by a clock) and distance (e.g.~of a physical object as measured by a suitable rod). It should be clear that this ``concretisation" of spacetime implies that its observed features and symmetries will be always related to properties and symmetries of the clocks and rods used to determine durations and distances and of the measurement protocols adopted.
The introduction of a set of observers endowed with standard clocks and rods and the agreement on a suitable procedure of synchronisation of such clocks  allows to rigorously define such a spacetime and observations tells us that locally (when gravity can be safely neglected) space is isotropic and space-time are homogeneous. Typically the notion of time as measured by our clocks is taken for simplicity in such a way that the motion of an isolated particle is uniform $\ddot{x} =0$. A system of observers satisfying this request is said to be inertial. One can then prove that if ${\cal K}$ is an inertial system of reference then any other system $\bar{\cal K}$ moving with respect to it of uniform rectilinear motion is inertial.

The above premises lay down the basic elements for an axiomatic derivation of the Lorentz  transformations~\cite{levy1976one,levy1979additivite,2005prst.book.....B}, here we shall discuss it following the logic proposed in~\cite{Sonego:2008iu}. Let's summarise the necessary assumptions.
\begin{itemize}
\item Spatial and temporal homogeneity
\item Spatial isotropy
\item The Relativity Principle (equivalence of inertial frames)
\item Pre-Causality (the time-ordering of two events along an observer worldline does not change in different systems of reference)
\end{itemize}

Let us start by considering two inertial frames ${\cal K} (O,x,y,z,t)$ and $\bar{\cal K} (\bar{O},\bar{x},\bar{y},\bar{z},\bar{t})$ where $O$ and $\bar{O}$ are respectively the origins of the two frames and we shall assume that ${\cal K}$ moves w.r.t.~$\bar{\cal K}$ with a uniform rectilinear speed $v$ (which, for simplicity, we can take along the $x$ direction). Note that a priori one can only presume that $v\in(-c_{-},c_{+})=J$ with $c_{-},c_+>0$ but possibly infinite. Similarly one can say that $\bar{\cal K}$ moves in ${\cal K}$ with velocity $\bar{v}$ (for the moment let us not assume anything about the $v$,$\bar{v}$ relation which we shall discuss later).

We can always choose the two origins to coincide at $t=\bar{t}=0$ i.e. $t=x=y=z=0$ if $\bar{t}=\bar{x}=\bar{y}=\bar{z}=0$ and we shall derive now what are the transformation laws relating the $\mathbf{x}$ and $\bar{\mathbf x}$ coordinates, $\bar{x}^\mu=f^\mu(t,x,y,z,v)$.

Now homogeneity in space and time requires that the above mentioned transformation laws are linear. This can be easily understood via the following reasoning: homogeneity implies that time or space intervals measured say in $\bar{\cal K}$ can depend only on the corresponding intervals in ${\cal K}$, not on the precise instants and positions when and where the measurements are made. 
Hence
\begin{equation}
{\rm d}\bar{x}^\mu=\frac{\partial f^\mu}{\partial x^\nu} {\rm d}x^\nu \qquad\mbox{with} \qquad \frac{\partial f^\mu}{\partial x^\nu}=\mbox{const}\; .
\end{equation}
I.e. $f^\mu$ must be a linear function of the coordinates $x^\mu$. Specialising for simplicity to the 1+1 case, the above statement translates in the simple form for the $(t,x)\to(\bar{t},\bar{x})$ transformations 
\begin{eqnarray}
\bar{t}&=&A(v)t+B(v)x\; ,\nonumber\\
\bar{x}&=&C(v)t+D(v) x\; .
\label{lin-tra}
\end{eqnarray}

Let us now ask how the motion of the origin $O$ appears in the system of reference $\bar{\cal K}$. We chose coordinates so that at $\bar{t}=0$ the origin $O$ had to be in $\bar{x}=0$, hence at any later time $\bar{x}_{O}=v\bar{t}$. However, we do also know that at any time $t$, $O$ has to be located at $x=0$ hence from (\ref{lin-tra}) we get for the coordinates of an event that takes place at $O$
\begin{eqnarray}
\bar{t}&=&A(v)t\; ,\nonumber\\
\bar{x}_{O}&=&C(v)t\; ,
\end{eqnarray}
from which (using $\bar{x}_{O}=v\bar{t}$) easily follows $C(v)=vA(v)$.

We can now repeat the same argument in describing the motion of $\bar{O}$ w.r.t.~${\cal K}$. This will move with speed $\bar{v}$ in ${\cal K}$ and reasoning as before we have ${x}_{\bar{O}}=\bar{v}t$ and given that $\bar{O}$ is always located in $\bar{x}=0$ in $\bar{\cal K}$ from (\ref{lin-tra}) we get
\begin{equation}
C(v)t+D(v)x_{\bar{O}}=0\;,
\end{equation}
from which (using ${x}_{\bar{O}}=\bar{v}t$) easily follows  $C(v)=-\bar{v} D(v)$ which, given the previous result, implies
\begin{eqnarray}
D(v)&=&-\frac{C(v)}{\bar{v}}=-\frac{v}{\bar{v}}A(v)\; .
\end{eqnarray}

We can then express the general transformation relating the two systems of reference as 
\begin{equation}
\left(\begin{array}{c}
\bar{t}\\ \bar{x}
\end{array}\right)=\Lambda(v)
\left(\begin{array}{c}
t\\ x
\end{array}\right)\;,
\label{trans-matrix}
\end{equation}
where $\Lambda(v)$ is the matrix
\begin{equation}
\Lambda(v):=A(v)
\left(\begin{array}{cc}
1 & \xi(v) \\
v & -{\textstyle v}/{\bar{\textstyle v}}
\end{array}\right)\;,
\label{lambda}
\end{equation}
and we have introduced the ratio $\xi(v):=B(v)/A(v)$.  

We can now make use of the isotropy of space (see \cite{Sonego:2008iu} for a more general derivation not assuming a priori isotropy) to relate $v$ and the reciprocal velocity $\bar{v}$. These are obviously related by the Relativity Principle which implies that if $v=F(\bar{v})$ then $\bar{v}=F(v)$ or equivalently $F=F^{-1}$. But this fact per se does not imply $\bar{v}=-v$, the so called Reciprocity Principle, as is normally postulated.
(Note that actually any $F$ symmetric under diagonal reflections $x\to-x$ would be Ok.) 
Indeed, one needs in addition to make explicit use of the isotropy of space requiring the symmetry of the transformation under an inversion of the spatial axis $x$~\cite{Berzi:1969bv,Torrettibook}. (It is interesting to note that von Ignatovski erroneously thought that the reciprocity principle can be derived as a consequence of the Relativity Principle.)

Hence, in conclusion, Eq.~(\ref{lambda}) further simplify into 
\begin{equation}
\Lambda(v):=A(v)
\left(\begin{array}{cc}
1 & \xi(v) \\
v & 1
\end{array}\right)\;,
\label{lambda2}
\end{equation}
Obviously, $\Lambda(0)$ must be the identity matrix, so $A(0)=1$ and $\xi(0)=0$. We can also check that invariance of the transformation under spatial inversion (e.g. $x\to-x$, $\bar{x}\to -x$, $u\to-u$) or time reversal requires $A(v)=A(-v)$ and $\xi(v)=-\xi(-v)$.

We are now ready to enforce the group structure into our set of transformation by appealing to the Relativity Principle.
The way to do so is to introduce a third system of reference $\bar{\bar{\cal K}}$ which moves with constant velocity $u$ w.r.t.~${\cal K}$ and ${\bar{u}}$ w.r.t.~$\bar{\cal K}$. We expect ${\bar{u}}$ to be given by some composition law 
\begin{equation}
{\bar{u}}=\Phi(u,v)\; ,
\end{equation}
while the transformation relating $\bar{\cal K}$ to $\bar{\bar{\cal K}}$ will be
\begin{equation}
\left(\begin{array}{c}
\bar{t}\\ \bar{x}
\end{array}\right)=\Lambda(\bar{u})
\left(\begin{array}{c}
\bar{\bar{t}}\\ \bar{\bar{x}}
\end{array}\right)\;.
\label{trans-matrix2}
\end{equation}
This, together with Eq.~(\ref{trans-matrix}), also tells us that the transformation from ${\cal K}$ to $\bar{\bar{\cal K}}$ will have the form
\begin{equation}
\left(\begin{array}{c}
{t}\\ {x}
\end{array}\right)=\Lambda(v)^{-1} \Lambda(\bar{u})
\left(\begin{array}{c}
\bar{\bar{t}}\\ \bar{\bar{x}}
\end{array}\right)\;.
\label{trans-matrix3}
\end{equation}
However, we also know that the transformation
\begin{equation}
\left(\begin{array}{c}
{t}\\ {x}
\end{array}\right)=\Lambda({u})
\left(\begin{array}{c}
\bar{\bar{t}}\\ \bar{\bar{x}}
\end{array}\right)\;.
\label{trans-matrix4}
\end{equation}
must independently hold which is tantamount to say that
\begin{equation}
\Lambda(v) ^{-1}\Lambda(\bar{u})=\Lambda(u)\quad\Longrightarrow\quad \Lambda(v)\Lambda(u)=\Lambda\left(\Phi(u,v)\right)\;.
\label{eq:group}
\end{equation}
This equation implies the group structure for our set of coordinate transformations. 
Note that if $u^{*}$ is some value of the velocity $u$ so that $\bar{u}=\Phi(u^{*},v)=0$ then $\Lambda(v)\Lambda(u^{*})= \Lambda(0)=\mathbb{I}$.	

Let us now differentiate w.r.t.~$v$ equation (\ref{eq:group}) and calculate it in $v=0$. We get 
\begin{equation}
\Lambda'(0)\Lambda(u)=\varphi(u) \Lambda'(u) \quad\mbox{where}\quad \varphi(u)=\left.\frac{\partial\Phi(u,v)}{\partial v}\right|_{v=0}\;.
\label{eq:group2}
\end{equation}
It is easy to see that this equation admits the solution
\begin{equation}
\Lambda(u)=\ee^{h(u)\Lambda'(0)}\quad\mbox{where}\quad h(u)=\int_0^u{\frac{{\rm d} \tilde{u}}{\varphi(\tilde{u})}}\; .
\label{expLambda}
\end{equation}
Now we can use Eq.~(\ref{lambda2}) to compute
\begin{equation}
\Lambda'(0)=A'(0) \mathbb{I}+\mathbb{M} \quad\mbox{where}\quad \mathbb{M}\equiv \left(\begin{array}{cc} 0 & \xi'(0)\\ 1 & 0 \end{array} \right)\; .
\end{equation}
The matrix $\mathbb{M}$ has the following algebraic properties:
For any natural number $k$,
\begin{equation}
\mathbb{M}^{2k}=\left(\xi'(0)\right)^k \mathbb{I}\;,\qquad\qquad
\mathbb{M}^{2k+1}=\left(\xi'(0)\right)^k \mathbb{M}\;.
\label{propM}
\end{equation}
Replacing these into Eq.\ (\ref{expLambda}) we find
\begin{equation}
\Lambda(u)=\ee^{h(u)A'(0)}
\left(\begin{array}{cc}
s_1(u) & \xi'(0)\,s_2(u) \\
s_2(u) & s_1(u)
\end{array}\right)\;,
\label{Lambda!}
\end{equation}
where:
\begin{equation}
s_1(u):=\sum_{k=0}^{+\infty}\frac{1}{(2k)!}\left(\xi'(0)\right)^k
h(u)^{2k}\;;
\label{sigma1}
\end{equation}
\begin{equation}
s_2(u):=\sum_{k=0}^{+\infty}\frac{1}{(2k+1)!}\left(\xi'(0)\right)^k
h(u)^{2k+1}\;.
\label{sigma2}
\end{equation}
We can now compare Eqs.\ (\ref{lambda2}) (with $v$ replaced by $u$) and (\ref{Lambda!}) to find:
\begin{equation}
A(u)=\ee^{h(u)A'(0)}s_1(u)\;;
\label{A}
\end{equation}
\begin{equation}
u=\frac{s_2(u)}{s_1(u)}\;;
\label{u}
\end{equation}
\begin{equation}
\xi(u)=\xi'(0)\,u\;;
\label{xi}
\end{equation}
Note that Eq.\ (\ref{u}) contains implicitly the link between $u$
and $h(u)$, because its right-hand side depends on $u$ only through $h(u)$.
Furthermore, we previously saw that $A(u)=A(-u)$ due to spatial or time reversal invariance. However, the function $h(u)$ is odd under this operation, $h(-u)=-h(u)$, while $s_1(u)$ is evidently even. By consistency one then must ask $A'(0)=0$ which in the end implies
\begin{equation}
\Lambda(u)=s_1(u) 
\left(\begin{array}{cc}
1 & \xi'(0)\,u \\
u & 1
\end{array}\right)\;,
\label{Lambda!2}
\end{equation}

Replacing Eq.\ (\ref{Lambda!2}) and the corresponding expressions for
$v$ and $\Phi(u,v)$ into Eq.\ (\ref{eq:group}), we find:
\begin{equation}
h(\Phi(u,v))=h(u)+h(v)\;;
\label{h+h}
\end{equation}
\begin{equation}
s_1(\Phi(u,v))=s_1(u)s_1(v)\left(1+\xi'(0)\right)\;;
\label{sigma1Phi}
\end{equation}
and most importantly
\begin{equation}
\Phi(u,v)=\frac{u+v}{1+\xi'(0)uv}\;.
\label{Phicomp}
\end{equation}
Equation (\ref{Phicomp}) implies  $\varphi(u)=-\xi'(0)\,u^2+1$, furthermore tells us that that the condition $\xi'(0)\geq0$ must be satisfied in order to guarantee that $\Phi(u,v)$ does not diverge for any finite value of $u$ or $v$. 

Finally, using the relationship $s_2'(u)=s_1(u)/\varphi(u)$ (which
follows from Eqs.\ (\ref{expLambda}), (\ref{sigma1}), (\ref{sigma2}) and (\ref{propM})) together with
Eqs.\ (\ref{u}) and $\varphi(u)=-\xi'(0)\,u^2+1$, we obtain
\begin{equation}
\frac{s_2'(u)}{s_2(u)}=\left(\frac{1}{u}\right)\frac{1}{-\xi'(0)\,u^2+1}=\frac{1}{u}
-\frac{\varphi'(u)}{2\,\varphi(u)}\;.
\label{sigma'/sigma}
\end{equation}
This can be immediately integrated to obtain, in a neighbourhood of
the origin, $s_2(u)=\pm u/\sqrt{\varphi(u)}$; hence $s_1(u)=\pm /\sqrt{\varphi(u)}$ and
\begin{equation}
A(u)=\frac{1}{\sqrt{\varphi(u)}}=\frac{1}{\sqrt{1-\xi'(0)u^2}}\;,
\label{A!}
\end{equation}
where a multiplicative constant and the sign ambiguity have been
fixed by the condition $A(0)=1$.  

All we need now is to impose what we called the pre-causality axiom which formally translates into the condition $\partial\bar{t}/\partial{t}>0$. In our set of transformations (\ref{trans-matrix}) this implies $A(u)>0\quad\forall u\in J$. However, we already know that $\xi'(0)\geq 0$, hence for $A(u)$ to be positive and real one has to require $\xi'(0)u^2<1$.

We can then define a speed scale $C^2\equiv 1/\xi'(0)$ and require $C^2>u^2$ for any $u$ and we can now write
\begin{equation}
\gamma(u)\equiv A(u)=\frac{1}{\sqrt{1-{ u^2}/{C^2}}}\; .
\end{equation}
The speed scale $C$ has two crucial properties
\begin{enumerate}
\item From (\ref{Phicomp}) we can easily see that it is a fixed point in the composition law $\Phi(C,v)=\Phi(u,C)=C$.
\item The transformations (\ref{trans-matrix}) holds only for $|u|<C$ so that $J=(-C,C)$.
\end{enumerate}
Note that we have not proved at all that signals cannot move faster than $C$! In principle these particles/signal are not incompatible with Lorentz transformations and they do not a priori lead to causality violations \cite{Liberati:2001sd}.

Finally our transformations (\ref{trans-matrix}) will then take the familiar form 
\begin{equation}
\Lambda(v):=\gamma(v)
\left(\begin{array}{cc}
1 & {v}/{C^2} \\
v & 1
\end{array}\right)\;,
\label{lambda-fin}
\end{equation}

It is only empirical observation which will tell us in the end if $C$ is finite (Lorentzian relativity) or infinite (Galilean relativity). Furthermore, nowhere in our derivation of the Lorentz transformations we needed to appeal to the universality of the speed of light. Even worse we do not know a priori that the limit speed $C$ has to coincide with such a physical entity. Let us stress that the determination of $C$ is strictly related to the physics underlying our system of clocks and rods. This was for example nicely described in a pedagogical novel by Trautman \cite{TrautBats} where bats using only sound wave to describe their spacetime would easily end up deriving the speed of sound as such an invariant/limit speed (see also \cite{Barcelo:2007iu} for an acoustic gravity analog concrete realisation of the same idea). 

In this sense the very notion of Lorentz invariance is detached by an abstract geometric entity as the Minkowski spacetime and firmly attached to a symmetry of the physics underling our tools for operationally define the spacetime. Conversely the Minkowski spacetime can be seen as a mathematical tool that summarises such properties.
In this sense tests of departures from Lorentz invariance can of course be seen on the one side as probes of the microscopic structure of space-time but can as well be considered high precision tests of the physics underlying our most precise clocks and rods.

\section{A historical overview}
\label{sec:hist}

The previous discussion should have convinced the reader about the fundamental nature and insight provided by Lorentz  invariance in modern physics. It is then not surprising that questioning and investigations regarding such fundamental symmetry appeared relatively early in last century literature. It is however only in recent times that speculations could be translated in full-fledged phenomenological studies. We shall here present an incomplete review of these developments.


The possibility that Lorentz invariance violation (LIV) could play a role again in physics dates back by at least sixty years~\cite{DiracAet,Bj,Phillips66,Blokh66,Pavl67,Redei67} and in the seventies and eighties there was already a well established  literature investigating how LI could be established at low energies without being an exact symmetry at all scales (see e.g.~\cite{1978NuPhB.141..153N,1980NuPhB.176...61E,Zee:1981sy,Nielsen:1982kx,1983NuPhB.217..125C,1983NuPhB.211..269N}). This stream of paper had a lasting influence but somehow remained as a set of isolated attempts to go beyond the standard framework.  In general, the gravitation theory community did not pay very much attention to the subject given
the general expectation at the time that possible breakdowns of standard physics due to quantum gravitational effects were only to appear in particle interactions at energies of the order Planck mass $M_{\rm Pl}$.  It was only in the nineties that it was clearly realised that there are special situations in which new effects could manifest also at lower energy. These situations were termed ``Windows on Quantum Gravity''.


While the Planck mass stands off from any standard model or observational scale (Ultra High Energy Cosmic Rays, see, e.g.,~\cite{Roth:2007in,Abbasi:2007sv}) have $E \lesssim 10^{11}$ GeV $\sim 10^{-8} M_{\mathrm Pl}$, sometimes even tiny deviations from standard physics can be magnified into a significant effect when dealing with threshold reactions at high energies (but still well below the Planck scale), long distances of signal propagation, or other very sensitive phenomena (see, e.g.,~\cite{Mattingly:2005re} for an extensive review).
A partial list of these {\em windows on QG} includes:
\begin{itemize}
\item sidereal variation of LIV couplings as the lab moves
  with respect to a preferred frame or direction
\item cumulative effects: long baseline dispersion and vacuum birefringence (e.g.~of signals from gamma ray bursts, active galactic nuclei, pulsars)
\item anomalous (normally forbidden) threshold reactions allowed by LIV terms (e.g.~photon decay, vacuum Cherenkov effect) 
\item shifting of existing threshold reactions (e.g.~photon annihilation from Blazars, ultra high energy protons pion production)
\item LIV induced decays not characterised by a threshold (e.g.~decay of a particle from one helicity to the other or photon splitting)
\item maximum velocity (e.g.~synchrotron peak from supernova remnants)
\item dynamical effects of LIV background fields (e.g. gravitational coupling and additional wave modes)
\end{itemize}

As usual in science, the realisation of this potentiality was a community effort, however one can identify some seminal papers that contributed to the take off of this field of research. Among these papers one can cite \cite{KS89} that already in 1989 envisaged, within a string field theory framework, the possibility of non-zero vacuum expectation values (VEV) for some Lorentz breaking operators. This work led later on to the development of a systematic extension of the SM (that was later on called "minimal standard model extension" (mSME)) incorporating all possible Lorentz breaking, gauge invariant, power counting renormalizable (i.e. of mass dimension $\leq 4$)~\cite{Colladay:1998fq}.  This provided a framework for computing in effective field theory the observable consequences for many experiments and led to much experimental work setting limits on the LIV parameters in the Lagrangian (see e.g.~\cite{Kostelecky:2008zz}).
 
Meanwhile, other theoretical investigations provided new motivations for Lorentz breaking searches and constraints. Indeed, specific hints of LIV arose from various approaches to Quantum Gravity. Among the many examples are the above mentioned string theory tensor VEVs \cite{KS89} and space-time foam models~\cite{AmelinoCamelia:1996pj,AmelinoCamelia:1997gz,Ellis:1999jf,Ellis:2000sx,Ellis:2003sd}. It was this stream of research that finally lead to the realisation that constraints could be cast also on high energy violations of Lorentz invariance in the photon dispersion relation, using the aforementioned propagation over cosmological distance of light from remote astrophysical sources like gamma ray bursters (GRBs) and active galactic nuclei (AGN), was instead highlighted in~\cite{AmelinoCamelia:1997gz}  The field of phenomenological constraints on quantum gravity induced LIV was born. 

In fact, starting from this early work a stream of theoretical and phenomenological papers followed. For example, the influential papers by Coleman and Glashow \cite{Coleman:1997xq,Coleman:1998en,Coleman:1998ti} which brought the subject of systematic tests of Lorentz violation to the attention of of the broader community of particle physicists. 
Let me stress, that several other papers appeared in the same period, some of them anticipating many important results, see e.g.~\cite{GonzalezMestres:1996zv,GonzalezMestres:1997if}. However at the time of their appearance were hardly noticed (probably seen by many as too ``exotic").

In the very same period an important part of the revived interest LIV was due to the development of Analogue Gravity~\cite{Barcelo:2005fc}. While the field was initiated in 1981 by Bill Unruh \cite{Unruh:1980cg} it was only in the early 2000 that it was considered more systematically as an inspirational source for possible new effects associated to an ``emergent gravity'' framework. In particular, modified dispersion relations of the sort suggested in these analogue systems provided an ideal setting for possible testable signature on the cosmic microwave background of Planck physics~\cite{Brandenberger:2000wr,Niemeyer:2001qe,Niemeyer:2002kh}. 

The above field was joined by many other suggestions from novel QG models/calculations. For example, semiclassical spin-network calculations in Loop QG~\cite{Gambini:1998it}, non-commutative geometry~\cite{Carroll:2001ws, Lukierski:1993wx, AmelinoCamelia:1999pm}, some brane-world backgrounds~\cite{Burgess:2002tb}. Furthermore, during the last three decades there were several attempts to formulate alternative theories of gravitation incorporating some form of Lorentz breaking, from early studies~\cite{Gasperini:1985aw,Gasperini:1986xb,Gasperini:1987nq,Gasperini:1987fq,Gasperini:1998eb} to full-fledged theories such as the Einstein--Aether theory \cite{Mattingly:2001yd,Eling:2004dk,Jacobson:2008aj} and Ho\v rava--Lifshitz  gravity \cite{Horava:2009uw,Sotiriou:2009bx,Blas:2009qj} (which in some limit can be seen as a UV completion of the Einstein--Aether framework~\cite{Jacobson:2010mx}). 

Coming back to the phenomenology side of this story, it was again in the early years 2000 that several works made an effort to provide a more systematic study both of LIV frameworks as well as of the available constraints (see e.g.~\cite{Jacobson:2002hd, Mattingly:2002ba, Jacobson:2005bg}). In this sense a crucial contribution was the development of an effective field theory approach also for higher order (mass dimension greater than four), naively non-power counting renormalizable, operators.~\footnote{Anisotropic scaling \cite{Anselmi:2008ry,Horava:2009uw,Visser:2009ys} techniques were recently recognised to be the most appropriate way of handling higher order operators in Lorentz breaking theories and in this case the highest order operators are indeed crucial in making the theory power counting renormalizable. This is why we shall adopt sometime the expression ``naively non renormalizable".} This was firstly done for dimension 5 operators in QED \cite{Myers:2003fd} to be later on extended to dimension 6 operators~\cite{Mattingly:2008pw}. 

Noticeably, in the very same period was advanced the idea that new relativity groups might exists  while being able to incorporate an invariant length/energy scale to be identified with the Planck one. This is the stream of research associated to the so called Doubly/Deformed Special Relativity~\cite{AmelinoCamelia:2000mn, Magueijo:2001cr,Magueijo:2002am, AmelinoCamelia:2002gv}). This field lead, quite recently, to more bold proposals such as Relative Locality \cite{AmelinoCamelia:2011bm,AmelinoCamelia:2011pe,Carmona:2011wc,AmelinoCamelia:2011yi,KowalskiGlikman:2012ji}. We shall here preeminently discuss tests of Lorentz violations within an EFT framework but we shall come back on these alternative framework in the end of this work.

It should be stressed that this burst of activity was not only spurred by the simple realisation that some phenomenological constraints on Planck physics could be cast. It is in fact important to note that the very same period was characterised by a growing plethora of observational puzzles seemingly hinting towards new physics. For example, in cosmology these are the years of the striking realization that our universe is undergoing an accelerated expansion phase \cite{Riess:1998cb,Perlmutter:1998np} which apparently requires a new exotic cosmological fluid, called dark energy, which violates the strong energy condition (to be added to the already well known, and still mysterious, dark matter component).

Also in the same period high energy astrophysics provided some new puzzles, first with the apparent absence of the Greisen-Zatsepin-Kuz'min (GZK) cut off \cite{Greisen:1966jv,1969cora...11...45Z} (a suppression of the high-energy tail of the UHECR spectrum due to UHECR interaction with CMB photons, see section \ref{sec:n4constr} below for further discussion) as claimed by the Japanese experiment AGASA \cite{Takeda:1998ps}, later on with the so called TeV-gamma rays crisis, i.e.~the apparent detection of a reduced absorption of TeV gamma rays emitted by AGN \cite{Protheroe:2000hp}. Both these ``crises" later on seemed to subside or at least alternative, more orthodox,  explanations for them were advanced. However, their undoubtedly boosted the research in the field at that time and recent observational developments leave these two puzzles still open to further scrutiny.

In this sense we cannot omit here the recent ``Opera affaire", i.e. the false detection by the CERN--LNGS based experiment OPERA~\cite{Opera:2011zb} of superluminal propagation of muonic neutrinos and Lorentz EFT (see e.g.~\cite{AmelinoCamelia:2011dx,Cohen:2011hx,Maccione:2011fr,Carmona:2011zg,Antonello:2012hg,Antonello:2012be}). While it is nowadays clear that the measurement was flawed due to unaccounted experimental errors, it did play an important role in propelling further activity in Lorentz breaking phenomenology which provided useful insights for future searches. We shall come back on this issue later using this case as a test field of the general techniques developed within a EFT framework.

\section{Matter Lorentz breaking effective field theory}
\label{sec:mattLIV}

Let us now discuss in detail the specific frameworks introduced within the EFT approach to Lorentz breaking phenomenology, which in the case of the matter sector is sometime referred to as the Standard Model Extension (SME). This consists in supplementing the standard model of particle physics with all the possible Lorentz violating operators that can be written without changing the field content or violating gauge symmetry. The operators appearing in the SME can be conveniently classified according to their mass dimension and behaviour under CPT. Note that Lorentz violation does not imply CPT violation for local EFTs~\cite{Chaichian:2011fc}, while CPT violation does imply Lorentz violation in local EFT~\cite{Greenberg:2002uu}.

Within the SME it is often considered the specific subset of the rotationally invariant LIV operators. One reason for restricting our analysis to this specific subgroup of operators is of course simplicity and clarity. In the full rotation breaking theory there appears very many independent (and often dimensionfull) coefficients which in turn lead to long tables of constraints from which it is sometime difficult to extract a sound physical intuition. Also theoretical arguments for LIV generally imply that it may arise in QG from the presence of a short distance
cutoff which suggests just breaking of boost invariance, with a preferred rest frame.\footnote{See however~\cite{Dowker:2003hb} for an example where (coarse grained) boost invariance is preserved in a
discrete model, and~\cite{Rovelli:2002vp,Livine:2004xy} for a study of how discreteness may be compatible with Lorentz symmetry in a quantum setting.} Furthermore, it would be very difficult for a theory which breaks rotation invariance to preserve boost
invariance -- if a spacelike four-vector is introduced to break rotation invariance, the four-vector also breaks boost invariance. Hence, a constraint on pure boost violation is, barring a conspiracy, also a constraint on boost plus rotation violation. We shall then from here on simplify our EFT treatment by always specialising to a rotational invariant setting.

Once rotational invariance is assumed, one can easily see that all LIV tensors must reduce to suitable products of a time-like vector field, usually denoted $u^{\alpha}$ which is usually taken to be unit (so that in the frame of the observer whose world line is tangent to $u^\alpha$, the latter has components $(1,0,0,0)$).   This allows us to express constraints solely in terms of the numerical coefficient involved in any $u^\alpha$-matter interaction term.  Of course, the actual direction of $u^\alpha$ is technically arbitrary.  However, the common choice, which we make here, is to define $u^\alpha$ to be aligned with the rest frame of the cosmic microwave background. Since the boost factor of an Earth centred frame is only $\sim10^{-3}$ \cite{Kogut:1993ag} with respect to the CMB frame, any rotation breaking effects generated by the Earth's relative motion with respect to the CMB will be naturally suppressed by a factor of $\sim10^{-3}$ relative to rotationally invariant effects and so we ignore them.

\subsection{mSME}

The Minimal Standard Model Extension or mSME, is the subset of the SME which considers only mass dimension 3 and 4 operators (naively power counting renormalizable operators).  For fermions and photons the rotationally invariant LIV operators then take the form
\begin{equation} 
\label{eq:LIVQEDelectronrotinv}
-a u_\mu \overline{\psi} \gamma^{\mu}\psi -bu_{\mu}\overline{\psi} \gamma_5 \gamma^{\mu}\psi + \frac {1} {2} i c
u_\mu u_\nu \overline{\psi} \gamma^{\mu}  \stackrel{\leftrightarrow}{D^{\nu}} \psi  + \frac {1} {2} i
d u_\mu u_\nu \overline{\psi} \gamma_5 \gamma^\mu  \stackrel{\leftrightarrow}{D^{\nu}} \psi \; ,
\end{equation}
\begin{equation}\label{eq:LIVQEDphotonrotinv}
-\frac{1}{4}(k_F){u_\kappa \eta_{\lambda\mu} u_\nu} F^{\kappa\lambda}F^{\mu\nu}+ \frac {1} {4} k_{AF} u^\kappa \epsilon_{\kappa \alpha \beta \gamma} A^\alpha F^{\beta \gamma}.
\end{equation}

The dimension 3, CPT odd $k_{AF}$ term generates an instability in the theory and so is generally set to zero.  The $a$ term can be absorbed by shifting the phase of the fermion field and so we will ignore it temporarily (note however that it can be relevant in gravity, as we shall see later, and in the case of flavour dependent LIV in neutrinos~\cite{Kostelecky:2011gq}).

The corresponding high energy ($M_{\rm Pl}\gg E \gg m$) dispersion relations for QED can be expressed as (see e.g.~\cite{Mattingly:2005re} and references therein)
\begin{eqnarray} \label{eq:SMErotinvdisp}
E_{\rm el}^2=m_{e}^2+p^2+f^{(1)}_e p+f^{(2)}_ep^2 \quad\mbox{electrons}\\
E_{\gamma}^2=(1+ f^{(2)}_\gamma ){p^2}\quad\mbox{photons}
\end{eqnarray}
where $f^{(1)}_e=-2bs,f^{(2)}_e=-(c-ds)$, and $f^{(2)}_\gamma=k_F/2$ with $s=\pm1$ the helicity state of the fermion~\cite{Mattingly:2005re}.
The anti-fermion dispersion relation is the same as (\ref{eq:SMErotinvdisp}) with the replacement $p\rightarrow -p$, which will change only the $f^{(1)}_e$ term.

Note that the typical energy at which new phenomenology should start to appear is quite low. In fact, taking for example $f_{e}^{(2)} \sim O(1)$, one finds that the corresponding extra-term is comparable to the particle mass $m$ precisely at $p \simeq m$. Even worse, for the linear modification to the dispersion relation, we would have, in the case in which $f^{(1)}_{e} \simeq O(1)$, that $p_{\rm th} \sim m^{2}/M_{\rm Pl} \sim 10^{-17}$~eV for electrons. 

\subsection{Rotational invariant CPT odd dimension 5 ops}
\label{sec:MP}

An alternative approach within EFT is to study naively non-renormalizable operators. It is nowadays widely accepted that the SM could just be an effective field theory and in this sense its renormalizability is seen as a consequence of neglecting some higher order operators which are suppressed by some appropriate mass scale (generally the Higgs or SUSY one). One can then speculate that at least some of these non-renormalizable operators can be generated by quantum gravity effects (and hence be naturally suppressed by the Planck mass), being associated to the violation of some fundamental space-time symmetry like local Lorentz invariance. 

In \cite{Myers:2003fd} it was found that there are essentially only three operators of dimension five, quadratic in the fields, that can be added to the QED\index{QED} Lagrangian preserving rotation and gauge invariance, but breaking local LI.~\footnote{Actually these criteria allow the addition of other (CPT even) terms, but these would not lead to modified dispersion relations (they can be thought of as extra, Planck suppressed, interaction terms) \cite{Bolokhov:2007yc}.}

These extra-terms, which result in a contribution of $O(E/M_{\rm Pl} )$ to the dispersion relation of the particles, are the following:
\begin{equation}
-\frac{\xi}{2M_{\rm Pl} }u^mF_{ma}(u\cdot\partial)(u_n\tilde{F}^{na}) + \frac{1}{2M_{\rm Pl} }u^m\overline{\psi}\gamma_m(\zeta_1+\zeta_2\gamma_5)(u\cdot\partial)^2\psi\:,
\label{eq:LIVterms}
\end{equation}
where $\tilde{F}$ is the dual of $F$ and $\xi$, $\zeta_{1,2}$ are dimensionless parameters. All these terms also violate the CPT symmetry.
More recently, this construction has been extended to the whole SM \cite{Bolokhov:2007yc}. 

From (\ref{eq:LIVterms}) the dispersion relations of the fields are
modified as follows. For the photon one has
\begin{equation}
\omega_{\pm}^2 = k^2 \pm \frac{\xi}{M_{\rm Pl} }k^3\:,
\label{eq:disp_rel_phot}
\end{equation}
(the $+$ and $-$ signs denote right and left circular polarisation), while
for the fermion (with the $+$ and $-$ signs now denoting positive and
negative helicity states%
)
\begin{equation}
E_\pm^2 = p^2 + m^2 + \eta_\pm \frac{p^3}{M_{\rm Pl} }\;,
\label{eq:disp_rel_ferm}
\end{equation}
with $\eta_\pm=2(\zeta_1\pm \zeta_2)$.  For the antifermion, it can be
shown by simple ``hole interpretation" arguments that the same
dispersion relation holds, with $\eta_{\overline{q}_\pm} = -\eta_{q_\mp}$ where
$\overline{q}$ and $q$ denote respectively anti-fermion and
fermion~\cite{Jacobson:2005bg,Jacobson:2003bn}.

As we shall see, observations involving very high energies can potentially cast $O(1)$ or stronger constraint on the coefficients defined above.  A natural question arises then: what is the theoretically expected value of the LIV coefficients in the modified dispersion relations shown above?  
This problem could be further exacerbated by renormalization group effects, which could, in principle, strongly suppress the low-energy values of the LIV coefficients even if they are $O(1)$ at high energies. Let us, therefore, consider the evolution of the LIV parameters first. This issue was addressed in~\cite{Bolokhov:2007yc} by calculating the renormalization group equations for QED and the Standard Model extended with dimension-five, CPT odd, operators. The running was found to be only logarithmic and therefore low energy constraints are robust: $O(1)$ parameters at the Planck scale are still $O(1)$ at lower energy. Moreover, they also show that $\eta_{+}$ and $\eta_{-}$ cannot, in general, be equal at all scales.
Similar results were found also in the mSME and are expected at all orders.

\subsection{Rotational invariant CPT even dimension 5 and 6 ops}
\label{sec:CPTev}

The CPT even, rotationally invariant mass dimension five and six LIV terms has been computed in~\cite{Mattingly:2008pw} using the same procedure described before for dimension 5 LIV. (This has been later extended to non-rotationally invariant operators in~\cite{Kostelecky:2009zp, Kostelecky:2011gq}.)  

With this ansatz one has for the photons
\begin{equation}
-\frac{1}{2M_{\rm Pl} ^{2}}\beta_{\gamma}^{(6)}F^{\mu\nu}u_{\mu}u^{\sigma}(u\cdot\partial)^2F_{\sigma\nu}\;.
\label{eq:op-dim6-phot}
\end{equation}
While for fermions the mass dimension five and six operators are
\begin{eqnarray}
-\frac{1} {M_{\rm Pl} } \overline{\psi} (u \cdot D)^2 (\alpha^{(5)}_L P_L + \alpha^{(5)}_R P_R) \psi\nonumber\\
 - \frac{i}{M_{\rm Pl} ^{2}}\overline{\psi}(u\cdot D)^{3}(u\cdot \gamma)(\alpha_{L}^{(6)}P_{L} + \alpha_{R}^{(6)}P_{R}) \psi \nonumber\\
-\frac{i}{M_{\rm Pl} ^{2}}\overline{\psi} (u\cdot D) \square (u\cdot \gamma) (\tilde{\alpha}_{L}^{(6)}P_{L} + \tilde{\alpha}_{R}^{(6)}P_{R}) \psi\;.
\label{eq:op-dim6-ferm}
\end{eqnarray}
From these operators, the dispersion relations of fermions and photons can be easily computed, yielding
\begin{eqnarray}
&&E^{2} - p^{2} - m^{2} =   \frac{\alpha_{R}^{(6)} E^{3}}{M_{\rm Pl} ^{2}}(E+sp) + \frac{\alpha_{L}^{(6)}E^{3}}{M_{\rm Pl} ^{2}}(E-sp)+ \nonumber\\
&&+\frac{m}{M_{\rm Pl} }(\alpha_{R}^{(5)}+\alpha_{L}^{(5)})p^{2} + \alpha_{R}^{(5)}\alpha_{L}^{(5)}\frac{p^{4}}{M_{\rm Pl} ^{2}}\label{eq:disp-rel-dimsix} \\
&&\omega^{2}-k^{2} = \beta^{(6)}\frac{k^{4}}{M_{\rm Pl} ^{2}}\;,
\label{eq:disp-rel-dimsix-phot} 
\end{eqnarray}
where $m$ is the electron mass and where $s = {\sigma}\cdot\mathbf{p}/|\mathbf{p}|$ is the helicity of the fermions, and we have neglected terms of  order $m^2 /M_{\rm Pl} ^2$ stemming from the $\tilde{\alpha}^{(6)}$ terms as they are highly suppressed. 

We can further simplify eq.~(\ref{eq:disp-rel-dimsix}) by noting that high-energy ($E\sim p$) fermions states are almost exactly chiral, and by grouping terms with same powers of the momentum. In this case we get  
\begin{equation}
E^{2} = p^{2} + m^{2}  + \frac{m} {M_{\rm Pl} } \eta^{(2)} p^{2} + \eta_{\pm}^{(4)} \frac{p^{4}}{M_{\rm Pl} ^{2}}\;
\label{eq:disp-rel-ferm-dim6-improved}
\end{equation}
where $R=+$, $L=-$ and we have labelled $\eta^{(n)}$ is the dispersion coefficient of the LIV $p^n$ term in the dispersion relation for the fermion.  We choose $\eta^{(n)}$ as the fermion coefficient symbol as this nomenclature is common in the literature.  Similarly, we shall use $\xi^{(n)}$ for the generic dispersion coefficient for a photon (so in (\ref{eq:disp-rel-dimsix-phot}) we shall take $\beta^{(6)}=\xi^{(4)}$).  

Let us briefly comment on the so obtained dispersion relations. Eq.~(\ref{eq:disp-rel-ferm-dim6-improved}) distinguishes different fermion helicities albeit being generated via the addition of CPT even terms, this is due to the presence of LIV terms among those displayed in eq.~(\ref{eq:op-dim6-ferm}) which are odd under P and T and hence can distinguish among helicity states.

In (\ref{eq:disp-rel-ferm-dim6-improved}) the quadratic modification generated by the dimension five operator is suppressed by a factor of order $m/M_{\rm Pl} $ and hence it can be safely neglected, provided that $E > \sqrt{mM_{\rm Pl} }$.  Note, however, that this term can be taken as an example of a dimension 4 LIV term with a natural suppression, which for electron is of order $m_{e}/M_{\rm Pl}  \sim 10^{-22}$. One should take this as a cautionary note as it shows clearly that for terms at this order, where no evident suppression is factorized out, the strength of a constraint should be judged only a posteriori, i.e.~on the base of a model predicting the natural size of the dimensionless coefficient in front of the $p^2$. In particular, to date the best constraint for a rotationally invariant electron LIV term of dimension 4 is $O(10^{-16})$ \cite{Stecker:2001vb} which would not be enough to effectively constrain the above mentioned term, given that its natural size is $10^{-22}$. 

By CPT, the dispersion relation of the anti-fermion is given by (\ref{eq:disp-rel-dimsix}), with the replacements $s \rightarrow -s$ and $p\rightarrow -p$. If $q,\overline{q}$ denote a charge fermion and anti-fermion, then the relevant anti-fermion coefficient $\eta^{(6)}_{\overline{q}}$ is such that $\eta^{(6)}_{\overline{q}_{\pm}} = \eta^{(6)}_{q_{\mp}}$, where $\overline{q}_{\pm}$ indicates an anti-fermion of positive/negative helicity (and similarly for the $q_{\pm}$). 
This should be compared with what we got from a  standard ``hole interpretation" argument in the case of CPT odd dimension five operators. It is easy to realise that this can be generalised via the same argument to conclude that for arbitrary ``n" order terms ($n$ being the power of the momentum) one would expect  $\eta^{(n)}_{\overline{q}_{\pm}} = (-1)^n \eta^{(n)}_{q_{\mp}}$. This different behaviour between even and odd powers of ``n" type dispersion relations leads to quite distinct phenomenologies as we shall see later.

\subsection{The Neutrino sector}
The generic neutrino LIV operators, at any mass dimension, have been categorized in \cite{Kostelecky:2011gq}. 
Also in this case, a significant reduction in the number of terms can be achieved by requiring that the LIV operators are rotationally symmetric.   Let us then focus on the Lagrangian for neutrinos with LIV operators of mass dimension up to six involving a vector field $u^{a}$ coupled to a Dirac neutrino $\psi$~
in a mass eigenstate with mass $m$ that are quadratic in matter fields and hence modify the free field equations.  We shall further assume that the Lorentz violation is diagonal in the mass basis. This might be justified by the idea that any theory of quantum gravity inducing such LIV must reduce to general relativity in the infrared, hence any Lorentz violation induced by quantum gravity would be primarily controlled by the charges that couple to gravity.  Of course, this does not meant that the coefficients for each mass eigenstate are the same, as RG effects would not allow them to be the same at any energy.

With these assumptions, the Lorentz violating terms (written in the mass basis) are exactly those for the QED fermions~\cite{Maccione:2011fr,Liberati:2012th}
\begin{eqnarray}
-a_i u_\mu \overline{\psi}_i \gamma^{\mu}\psi_i -b_i u_{\mu}\overline{\psi}_i \gamma_5 \gamma^{\mu}\psi_i + \frac {1} {2} i c_i u_\mu u_\nu \overline{\psi}_i \gamma^{\mu}  \stackrel{\leftrightarrow}{\partial^{\nu}} \psi_i  \\ \nonumber + \frac {1} {2} i
d_i u_\mu u_\nu \overline{\psi}_i \gamma_5 \gamma^\mu  \stackrel{\leftrightarrow}{\partial^{\nu}} \psi_i +\frac{1}{2\Mpl}u^m\overline{\psi}_i\gamma_m(\zeta_{i,1}+\zeta_{i,2}\gamma_5)(u\cdot\partial)^2\psi_i\\ \nonumber
 - \frac{i}{\Mpl^{2}}\overline{\psi}_i(u\cdot \partial)^{3}(u\cdot \gamma)(\alpha_{i,L}^{(6)}P_{L} + \alpha_{i,R}^{(6)}P_{R}) \psi_i
\end{eqnarray}
where $i$ is the mass index.  We've dropped the gauge covariant derivative above, as it is irrelevant and couples in the flavor basis so merely would add needless complication.   Also note that we have included both right and left projection operators, albeit, since standard model interactions only produce left-handed neutrinos, the constraints will primarily be on the corresponding left-handed operators.  Finally, in contrast to the QED case, one cannot drop the $a_i$ term here as this gives a contribution to the oscillation pattern.

The above terms and the usual Dirac Lagrangian for the neutrino yield a high energy neutrino dispersion relation of
\begin{eqnarray}
\label{eq:disp_rel_nu}
E_i^2 = p^2 + N_i^2 \\
\nonumber N_i^2=m_i^2 + 2(a_i+b_i)p-(c_i+d_i) p^2
 + 2(\zeta_{i,1}-\zeta_{i,2}) \frac{p^3}{\Mpl} + 2 \frac{\alpha_{i,L}^{(6)}p^{4}}{\Mpl^{2}} 
\end{eqnarray}
As before, the term proportional to $m/\Mpl$ has been ignored.  The net result of the LIV terms is to modify the dispersion relation of each mass eigenstate according to 
\begin{equation} 
E^2 - p^2 - (m_i)^2 = \sum_{n=1}^{4} \xi_i^{(n)} \frac{|p|^n}{\Mpl^{n-2}},
\label{eq:mdrnu}
\end{equation}
where $\xi_i^{(n)}$ is a coefficient that depends on the relevant terms in the Lagrangian~(\ref{eq:disp_rel_nu}), so that constraints on the $\xi_i^{(n)}$ can always be translated in constraints on the coefficients of Eq.~(\ref{eq:disp_rel_nu}). The corresponding anti-particle dispersion relation is easily derived by considering the behaviour of each term under CPT and given by
\begin{equation} 
E^2 - p^2 - (m_i)^2 = \sum_{n=1}^{4} (-1)^n \xi_i^{(n)} \frac{|p|^n}{\Mpl^{n-2}}.
\label{eq:antimdr}
\end{equation}
We leave the index $n$ as a free phenomenological parameter and consider the cases $n=2,3,4$ separately (the case $n=1$ would produce huge effects at low energy and is strongly constrained).  

Note also that many existing neutrino oscillation experiments measure the transition probability between neutrino flavours $P_{IJ}$ ($I$ and $J$ here labelling neutrino flavours) which is affected by the modified dispersion relation (\ref{eq:disp_rel_nu}). 
\begin{equation}
P_{IJ} = \delta_{IJ} - \sum_{i,j>i}4F_{IJij}\sin^2\left(\frac{\delta N_{ij}^2L}{4E}\right)+2G_{IJij}\sin^2\left(\frac{\delta N_{ij}^2L}{2E}\right)\;,
\end{equation}
with $\delta N_{ij}^2 = N_i^2-N_j^2$ and $F_{IJij}$ and $G_{IJij}$ are functions of the mixing matrixes.  Many of these experiments also quote results on a deviation of the neutrino speed from that of light, i.e.
\begin{equation}
\left(\frac{\Delta c}{c}\right)^{LIV}_{ij}= E^{-2}(\delta N_{ij}^2 - \delta m_{ij}^2)
\end{equation}
which can be easily translated into constraints on the coefficients of Eq.~(\ref{eq:disp_rel_nu}).

\section{Lorentz breaking and naturalness}

We presented in Section~\ref{sec:hist} an overview of several quantum gravity proposals/models which contributed in motivating the study of Lorentz breaking phenomenology.  A common feature of these proposals is the presence of  modified dispersion relations as those presented in the previous sections. For the rotational invariant case, these relations can be cast in the general form~\footnote{Note that we are not considering here dissipative terms in the dispersion relation, as they would require a separate treatment (e.g.~ with respect to the preservation of unitarity)  \cite{Parentani:2007uq}. Nonetheless, such dissipative scenarios are logically consistent and even plausible within some quantum/emergent gravity frameworks.}
%
%
\begin{equation}%
E^2=p^2+m^2+\sum_{n=1}^{\infty} \tilde{\eta}_{(n)} p^n\;,%
\label{eq:disprel}%
\end{equation}%
where the $\tilde{\eta}_i$ are now dimensional coefficients. This is also why we considered increasing mass dimension operators in EFT providing dispersion relations at some definite order in $n$.

However, while a definite QG model could predict different strengths for coefficients of different order (e.g. due too possible intrinsic symmetries of the model), from an EFT point of view the only relevant operators should be the lowest order ones, i.e. those of mass dimension 3,4 corresponding to terms of order $p$ and $p^2$ in the dispersion relation. Situations in which higher order operators ``weight" as much as the lowest order ones are only possible at the cost of a severe fine tuning of the coefficients $\tilde{\eta}_{(n)}$.  

The reason for this is pretty simple: in EFT radiative corrections will generically allow the percolation of higher dimension Lorentz violating terms into the lower dimension terms due to the interactions of particles~\cite{Collins:2004bp,Polchinski:2011za,Iengo:2009ix}.  Indeed,  EFT loop integrals will be naturally cut-off at the EFT breaking scale, if such scale is as well the Lorentz breaking scale the two will effectively cancel leading to unsuppressed, coupling dependent, contributions to the mass dimension four kinetic terms that generate the usual propagators.  Hence radiative corrections will not allow a dispersion relation with only $p^3$ or $p^4$ Lorentz breaking terms but will automatically induce extra unsuppressed LIV terms in $p$ and $p^2$ which will be naturally dominant.  One could argue that RG effects might naturally suppress the sizes of these coefficients at low energies.  However, in specific models where the RG flow has been calculated, the running of LIV coefficients is only logarithmic and so there is no indication that RG flow will actually drive coefficients to zero quickly in the infrared~\cite{Bolokhov:2007yc}.

Current observational constraints are tremendous on dimension 3 operators and very severe on dimension 4 ones. This is kind of obvious, as dimension 3 operator would dominate at $p\to 0$ while the dimension 4 ones would generically induce a, species dependent, constant shift in the limit speed for elementary particles. Hence, within a ``blind" EFT approach, it would deem unreasonable to further test departure from Lorentz invariance.

Nonetheless, missing a definitive QG prediction the phenomenological approach that has been taken in the past is to fix an order of $n$ and cast suitable constraints on the corresponding dispersion relation. This is is substantially equivalent to assuming a hierarchy  of LIV coefficients of the sort
\begin{equation}
\tilde{\eta}_1=\eta_1 \frac{\mu^2}{M},\qquad
\tilde{\eta}_2=\eta_2 \frac{\mu}{M},\qquad \tilde{\eta}_3=\eta_3
\frac{1}{M}
\label{disprel2} 
\end{equation}
or
\begin{equation}
 \tilde{\eta}_2=\eta_2 \frac{\mu^2}{M^2},\qquad
\tilde{\eta}_4=\eta_4 \frac{1}{M^2}
\label{disprelnoodd} 
\end{equation}
where $M$ is the Lorentz breaking scale, that we assumed so far to be equal to the Planck scale, $\mu$ is some other far lower energy scale, and $\eta_{(n)}$ is now the de-demensionalized coefficient usually assumed to be $O(1)$.  The exact hierarchy, whether it involves terms of every mass dimension as in (\ref{disprel2}), or only even dimension as in (\ref{disprelnoodd}) is model dependent.  

Several ideas have been advanced in order to justify such an unnatural (at least in EFT) hierarchy~\cite{Jacobson:2005bg}, sometime leading to a vibrant debate~\cite{Gambini:2011nx,Polchinski:2011za}, but one can clearly see that the most straightforward solution for this problem would consist in breaking the degeneracy between the EFT scale and the Lorentz breaking one.  We now briefly describe threes ideas that have been put forward that would generate such scales.

\subsection{A custodial symmetry?}

One (at least conceptually) simple solution, consists in appealing to some sort of  ``custodial symmetry", i.e.~a symmetry, other than Lorentz, which would forbid the appearance of the lower dimension operators and is broken at the low energy (EFT) scale $\mu$. The most natural symmetry that may play this role is of course supersymmetry (SUSY)~\cite{GrootNibbelink:2004za,Bolokhov:2005cj}. Albeit SUSY is a symmetry relating fermions to bosons  i.e.~matter with interaction carriers, it also as a matter of fact, intimately related to Lorentz invariance. Indeed, it can be shown that the composition of at least two SUSY transformations induces space-time translations. However, SUSY can still be an exact symmetry even in presence of LIV and as such it can actually play the role of a custodial symmetry. 

It can be shown that the effect of SUSY on LIV EFT is to prevent dimension $\leq 4$, LIV operators to be present in the classical Lagrangian. Furthermore, the renormalization group equations for Supersymmetric QED with dimension 5 LIV operators as discussed in section \ref{sec:MP} do not generate lower dimensional operators if SUSY is unbroken~\cite{GrootNibbelink:2004za,Bolokhov:2005cj}. 

Of course, SUSY is broken at low energies. The effect of soft SUSY breaking was also investigated in  
One should then expect, as it is confirmed by direct calculations \cite{GrootNibbelink:2004za,Bolokhov:2005cj}, that  SUSY breaking allows the percolation of high energy LIV to the renormalizable operators. In particular, dimension $\kappa$ ones arise from the percolation of dimension $\kappa+2$ LIV operators. (We consider here only $\kappa = 3,4$, for which these relationships have been demonstrated). However, in this case the coefficients appearing in front of the so generated low mass dimension operators will be naturally small being suppressed by powers of the ration of the SUSY breaking scale with the Planck scale. Typically the suppression is of order $m_{s}^{2}/M_{\rm Pl}$ ($\kappa=3$) or $(m_{s}/M_{\rm Pl})^{2}$ ($\kappa=4$), where $m_{s}$ is the scale of SUSY soft breaking. Although, given present constraints, the theory with $\kappa=3$ needs a lot of fine tuning to be viable, since the SUSY-breaking-induced suppression is not enough powerful to suppress enough linear modifications in the dispersion relation of electrons. 
Of course, one could always get rid e.g. in LIV QED of the terms of dimension 5 discussed in section~\ref{sec:MP} by just imposing CPT invariance as a further symmetry of the problem and consider the next order operators like in section \ref{sec:CPTev} . In this case one would end up inducing mass dimension 4 operators which are suppressed enough, provided $m_{s} < 100$~TeV. Current lower bounds from the Large Hadron Collider are at most around 950 GeV for the most simple models of SUSY~\cite{ATLAS} (the so called ``constrained minimal supersymmetric standard model", CMSSM).

As a final remark, let us notice that analogue models of gravity can be used as a particular implementation of the above mentioned mechanism for avoiding the so called naturalness problem via a custodial symmetry. This was indeed the case of multi-BEC~\cite{Liberati:2005pr,Liberati:2005id}.

\subsection{Gravitational confinement?}
\label{gravconf}

One possible alternative to the custodial symmetry idea consists in basically turning upside down the problem and assume that the Lorentz breaking scale is not set by the Planck scale.  If one does this and begins with a theory which has higher order Lorentz violating operators only in the gravitational sector (e.g. Ho\v rava--Lifshitz gravity), then one can hope that the gravitational coupling $G_N\sim M^{-2}_{\rm Pl}$ will suppress the ``percolation" to the matter sector where the constraints are strongest.  Matter Lorentz violating terms will all possess factors of the order $(\mu/M_{\rm Pl})^{2}$ which can become strong suppression factors if $\mu\ll M_{\rm Pl}$~\cite{Pospelov:2010mp}. Of course, this is done at the dear price of large LIV effects in the gravity sector. However, the constraints we have on high energy violations in gravity are very poor so far, and hence this mechanism cannot be ruled out at the moment. We shall come back on this proposal later on in this review.

\subsection{Improved RG flows by strong dynamics}
\label{RGflow}

One simple idea for solving the above mentioned naturalness problem is to construct a SME in such a way that the renormalization group flow of the LIV parameters (e.g. of $\delta c/c$ for elementary particles) will run to LI values in the IR.  This was basically the original idea behind the pioneering work by Nielsen and collaborators in the early eighties~\cite{1978NuPhB.141..153N,1980NuPhB.176...61E,Zee:1981sy,Nielsen:1982kx,1983NuPhB.217..125C,1983NuPhB.211..269N}. Unfortunately, as mentioned before, that program could never fulfil its scope because the LI values of the relevant quantities, albeit being attractors in the IR, they were so only with logarithmic flow, i.e. too slow for preventing large deviations at relative low energies. However, there was recently a resurgence of activity in the field~\cite{Sundrum:2011ic,Bednik:2013nxa}. The basic idea would be in this case to enhance at very high energies the RG flow via a strong coupled dynamics so that when approaching the logarithmic IR regime most of the suppression of LIV factors has already be achieved. Albeit preliminary, the results in this sense appear to be promising~\cite{Bednik:2013nxa}.

%

\section{A toolkit for testing of Lorentz invariance}

Tests of Lorentz invariance have an early history given that tests of Special Relativity have been performed, in increasing accurate ways, sine its initial formulation in 1905. However, it is way more recent, as we discussed in our historical review, the development of a systematic EFT-based approach to the problem. While these kind of blind searchers are not directly related to a specific QG approach and hence can at time be difficult to asses in their physical consequences, it was undoubtedly an important step tho categorisation of the possible departures from extant Lorentz symmetry as they allowed a more clear understanding of what was going to be constrained and what kind of observations/experiments had to be seek for in order to do so.

The phenomenological toolkit that was developed also thanks to this systematic approach is now quite rich and can be split in two big subsets: terrestrial experiments and astrophysical observations.

\subsection{Experimental probes of low energy LIV: Earth based experiments}
 
It is an observational fact that Nature as we probe it well below the Planck scale it is Lorentz invariant to a very high degree. It is hence logic that in seeking for low energy deviations from this symmetry, as those systematically described by the minimal Standard Model extension (mSME), one has to resort to very high precision experiments. We shall now briefly list the most exploded observations in this sense (for more details see e.g.~\cite{Mattingly:2005re,Kostelecky:2008zz,Kostelecky:2008ts}).

\subsubsection{Clock comparison Experiments}

Two co-local atomic transition frequencies can be considered as two clocks. As the clocks move in space, they pick out different components of the Lorentz violating tensors in the mSME. This in turn yields a sidereal drift between the two clocks which could be constrained by measuring the difference between the frequencies over long periods. This technique allows to cast very high precision limits on the parameters in the mSME (generally for for protons and neutrons). 

\subsubsection{Cavity Experiments}

The technique adopted in cavity experiments casts constraints on the variation of the cavity resonance frequency (with respect to a stationary frequency standard) as its orientation changes in space. While this is intrinsically similar to clock comparison experiments, these kind of experiments allows to cast constraints also on the electromagnetic sector of the mSME (as one of the ``clocks" in this case involves photons).

\subsubsection{Neutral mesons}

The mass difference of neutral mesons is one of the most accurately quantities in the SM.  The SME operators do affect such a quantity in a Lorentz breaking way, hence one generically expects an orientation dependent change which can be constrained by  looking for sidereal variations and other orientation effects. Also lifetime directional dependence has been investigated (see e.g.~\cite{DeAngelis:2010dg}).

\subsubsection{Penning traps}

In a Penning trap a combination of static magnetic and electric fields confines a charged particle for long times. One can then problem possible deviation from exact Lorentz invariance by monitoring the particle cyclotron motion in the magnetic field and its Larmor precession due to the spin. In fact the relevant frequencies for both these motions are affected by some mSME operators and Penning traps can be set up so to make them very sensitive to differences in these frequencies.

\subsubsection{Spin polarized torsion balance}

Spin-torsion balances are a very effective tool for constraining the electron sector of the mSME.
An example of such balances consists in an octagonal pattern of magnets which is constructed so to have an overall spin polarization in the octagonÕs plane. Four of these octagons are suspended from a torsion fiber in a vacuum chamber so to give an estimated net spin polarization equivalent to $\approx10^{23}$ aligned electron spins. In order to detect Lorentz breaking effect one has again to look for orientation dependent phenomena. For this reason the whole apparatus is mounted on a turntable. As the turntable moves, some of the Lorentz violation terms in the mSME, end up producing an interaction potential for non-relativistic electrons which induces a torque on the torsion balance. The torsion fiber is then twisted by an amount related to the relevant LIV coefficients which in this way can be constrained.

The absence of a preferred direction is also checked with great precision using nuclear spin, which translates into more stringent limits on LIV operators for the light quarks and photons (the photon LIV operator will also contribute because of the electromagnetic interactions inside the nucleon).
This possibility was analysed for the mSME as week as for the SME with dimension 5, CPT odd operators \cite{Myers:2003fd}. These experiments are again based on a search for orientation dependent phenomena using comagnetometers of different types \cite{Brown:2010dt,Smiciklas:2011xq} able to detect the sidereal variation with great accuracy. These methods are currently able to constraints the coefficients quarks and photons at $O(10^{-7}\div10^{-8})$. 

Remarkably, the constraints of $O(10^{-29})$ cast in \cite{Smiciklas:2011xq} on anisotropic components of the limit speed for neutrons in the mSME can also be converted in a severe constraint also on any time-like component (i.e.~rotational invariant). In fact, we assumed the preferred frame to be aligned with the CMB but the solar system is moving relative to such rest frame with boost factor $\sim 10^{?3}$ at a declination of $-7^\circ$. The missing alignment will hence always induce an anisotropic component suppressed by the square of the boost factor.
In conclusion, this allows to constrain an eventual isotropic component in the neutron limit speed at $O(10^{-23})$. 

\subsubsection{Recoil of slow atoms}

In alternative to the relativistic expansion that we have considered so far, one might as well consider non-relativistic limit, $p\ll m$, of  the phenomenological dispersion relations of the form (\ref{eq:disprel}) supplemented with the further assumption (\ref{disprel2}). It is easy to see that in such a case the dispersion relations and its most relevant correction would look like 
\begin{equation}
E\approx m+\frac{p^2}{2m}+\frac{1}{2M}\left(\eta_1 mp+\eta_2 p^2+\frac{p^3}{m}\right)\, .
\end{equation}
It was recently noticed~\cite{AmelinoCamelia:2009zzb,Mercati:2010au} that ultra-precise cold-atom-recoil experiments have now reached sufficient sensitivities for constraining the lowest coefficients in the above expansion (albeit not strongly). The basic techniques consists in using the available measurements of the ``recoil frequency" of atoms with experimental setups involving one or more ``two-photon Raman transitions"~\cite{Kasevich:1991zz,Peters99,Wicht02}. The basic idea is that one can give momentum to an atom through a process involving absorption
of a photon of frequency $\nu$ and a stimulated emission, in the opposite direction, of a photon of frequency $\nu'$. A straightforward application of energy-momentum conservation implies~\cite{AmelinoCamelia:2009zzb,Mercati:2010au}
\begin{equation}
     \frac{\Delta \nu}{ 2 \nu_* (\nu_* +p/h)} = \frac{h}{m} ~, 
     \label{coldat}
\end{equation}
where $\nu_*$ is resonance frequency  of the atom, $p$ is the momentum transferred to the atom and $\Delta \nu \equiv \nu - \nu'$.
The crucial point is that $h \Delta \nu \simeq  E(p + h\nu + h\nu') - E(p) \simeq E(2  h\nu_* + p) - E(p)$ and hence would be affected by modification of the dispersion relation, modifications which, thanks to the high precisions of current measurements could be then use to constraint the lowest order coefficients $\eta_1$ and $\eta_2$ ~\cite{AmelinoCamelia:2009zzb,Mercati:2010au}.

\subsection{Observational probes of high energy LIV: astrophysical QED reactions}
 
Testing the higher mass dimensions operators (5 or 6) of the SME is obviously a task that requires to probe much higher energies than those achieved in terrestrial experiments. Energies up to $10^{20}$~eV are achieved in high energy astrophysics and as such this field has played an eminent role in QG phenomenology. One comment is in order before we further describe these tests and it concerns our parametrisation. We have introduced dimensionless coefficients $\eta^{(n)}$ and expressed the higher order terms in the dispersion relations via suitable rations of the particle momentum to the Planck scale. Of course nothing guarantees that models of quantum/emergent spacetime will predict the Loretz breaking scale to be coincident with the Planck scale.  So, missing a derivation of our dispersion relation from a specific QG model, this has to be considered just a convenient parametrisation.

Let us begin with a brief review of the most common types of reaction exploited in order to give constraints on the QED sector.
For definiteness, we refer to the following modified dispersion relations:
\begin{eqnarray}
\label{eq:mdrqed}
E^{2}_{\gamma} &=& k^{2} + \xi_{\pm}^{(n)}\frac{k^{n}}{\Mpl^{n-2}}\qquad \mbox{Photon}\\
E^{2}_{el} &=& m_{e}^{2} + p^{2} + \eta_{\pm}^{(n)}\frac{p^{n}}{\Mpl^{n-2}}\qquad\mbox{Electron-Positron}\;,
\end{eqnarray}
where, in the EFT case, we have $\xi^{(n)} \equiv \xi_{+}^{(n)} = (-)^{n}\xi_{-}^{(n)}$ and $\eta^{(n)} \equiv \eta_{+}^{(n)} = (-)^{n}\eta_{-}^{(n)}$.

\subsubsection{Photon time of flight\index{Photon time of flight}}
\label{subsec:tof}

Photon time-of-flight were historically among the first tests proposed~\cite{AmelinoCamelia:1997gz} for QG phenomenology and although within an EFT framework they currently provide limits several orders of magnitude weaker than the best ones, they have been widely adopted in the astrophysical community.  Furthermore, being purely kinematical in nature, they apply on any modified dispersion relation independently from the underlying dynamical framework. Let us then discuss a general description of time-of-flight effects, elaborating later about their application to the specific case of EFT.

A photon dispersion relation in the form of (\ref{eq:mdrqed}) implies that photons of different colours (wave vectors $k_1$ and $k_2$) travel at slightly different speeds.  We shall assume here that no birefringent effects are present, so that $\xi_{+}^{(n)} = \xi_{-}^{(n)}$ (or alternatively in EFT we shall consider one helicity at a time). We propagating on a cosmological distance $d$, the effect of energy dependence of the photon group velocity will generically produces an energy dependent  time delay
\begin{equation}
 \Delta t^{(n)} = \frac{n-1}{2}\, \frac{k_2^{n-2}-k_1^{n-2}}{\Mpl^{n-2}}\,\xi^{(n)}\, d\;,
\label{eq:tof-naive}
\end{equation}
which increases linearly with $d$~\footnote{
Note that for an object located at cosmological distance (let $z$ be its redshift), the distance $d$ becomes
\begin{equation}
d(z) = \frac{1}{H_{0}}\int^{z}_0 \frac{1+z'}{\sqrt{\Omega_{\Lambda} + \Omega_{m}(1+z')^{3}}}\,dz'\;,
\end{equation}
where $d(z)$ is not exactly the distance of the object as it includes a $(1+z)^{2}$ factor in the integrand to take into account the redshift acting on the photon energies.}
and with the energy difference as long as $n>2$. The largest systematic error affecting this method is the uncertainty about whether photons of different energy are produced simultaneously in the source (generally Gamma-Ray Bursts or Active Galactic Nuclei).

So far, the most robust constraints on $\xi^{(3)}$, derived from time of flight differences, have been cast adopting a statistical analysis applied to the arrival times of sharp features in the intensity at different energies from a large sample of GRBs with known redshifts~\cite{Ellis:1999sd,Ellis:2005wr}, leading to limits $\xi^{(3)}\leq O(10^3)$.
A recent example illustrating the importance of systematic uncertainties can be found in \cite{Albert:2007qk}, where the strongest limit $\xi^{(3)} < 47$ is found by looking at a very strong flare in the TeV band of the AGN Markarian 501. Furthermore, analysing the flare time-structure, the authors also obtained a best fit for $\xi^{(3)} \sim O(1)$. However, a similar fit could also be achieved by standard plasma physics.

In order to alleviate systematic uncertainties  one could resort to extra information based on the specific dynamical model adopted. In particular in EFT one expects dimension 5 CPT odd terms to produce birefringent photon dispersion relation. One could then try to measure the velocity difference between the two polarization states at a single energy photon, corresponding to
\begin{equation}
 \Delta t = 2|\xi^{(3)}|k\, d/\Mpl\;.
\end{equation}
This bound would require that both polarizations be observed and that no spurious helicity-dependent mechanism (e.g.~propagation through a birefringent medium) affects the relative propagation of the two polarization states. As such this possibility has not been exploited so far.

More interestingly, Eq.~(\ref{eq:tof-naive}) is not necessarily valid in birefringent theories. In fact, photon beams generally are not circularly polarized; thus, they are a superposition of fast and slow modes. The net effect of this superposition may end-up erasing the time-delay effect. We can compute this effect on a generic photon beam in a birefringent theory by means of the associated electric field, and let us assume that this beam has been generated with a Gaussian width
\begin{equation}
\vec{E} = A\, \left(e^{i(\Omega_{0}t-k^{+}(\Omega_{0})z)}\,e^{-(z-v_{g}^{+}t)^{2}\delta\Omega_{0}^{2}}\hat{e}_{+} + e^{i(\Omega_{0}t-k^{-}(\Omega_{0})z)}\,e^{-(z-v_{g}^{-}t)^{2}\delta\Omega_{0}^{2}}\hat{e}_{-} \right)\, ,
\end{equation}
where $\Omega_{0}$ is the wave frequency, $\delta\Omega_{0}$ is the gaussian width of the wave, $k^{\pm}(\Omega_{0})$ is the ``momentum'' corresponding to the given frequency according to (\ref{eq:mdrqed}) and $\hat{e}_{\pm}\equiv (\hat{e}_{1}\pm i\hat{e}_{2})/\sqrt{2}$ are the helicity eigenstates. Note that by complex conjugation $\hat{e}_{+}^{*} = \hat{e}_{-}$. Also, note that $k^{\pm}(\omega) = \omega \mp \xi \omega^{2}/\Mpl$. 
Thus,
\begin{equation}
\vec{E} = A\, e^{i\Omega_{0}(t-z)}\left(e^{i\xi\Omega_{0}^{2}/\Mpl z}\,e^{-(z-v_{g}^{+}t)^{2}\delta\Omega_{0}^{2}}\hat{e}_{+} + e^{-i\xi\Omega_{0}^{2}/\Mpl z}\,e^{-(z-v_{g}^{-}t)^{2}\delta\Omega_{0}^{2}}\hat{e}_{-}       \right)\;.
\end{equation}
The intensity of the wave beam can be computed as
\begin{eqnarray}
\nonumber \vec{E}\cdot\vec{E}^{*} &=& |A|^{2}\left( e^{2i\xi\Omega_{0}^{2}/\Mpl z} + e^{-2i\xi\Omega_{0}^{2}/\Mpl z}  \right) e^{-\delta\Omega_{0}^{2}\left( (z-v_{g}^{+}t)^{2} + (z-v_{g}^{-}t)^{2}\right)}\\
&=& 2|A|^{2}e^{-2\delta\Omega_{0}^{2}(z-t)^{2}}\cos\left( 2\xi\frac{\Omega_{0}}{\Mpl}\Omega_{0}z\right)e^{ - 2\xi^{2}\frac{\Omega_{0}^{2}}{M^{2}}(\delta\Omega_{0}t)^{2}}\;.
\end{eqnarray}
We see that there is a modulation of the wave intensity that depends quadratically on the energy and linearly on the distance of propagation. In addition, for a Gaussian wave packet, there is a shift of the packet centre, that is controlled by the square of $\xi^{(3)}/\Mpl$ and hence is strongly suppressed with respect to the cosinusoidal modulation. So far, to the best of our knowledge, this EFT specific features have not been tested (and as we shall see below they would be extremely difficult to see anyway as birefringence is strongly constrained by current observations).

\subsubsection{Vacuum Birefringence\index{Vacuum Birefringence}}
\label{sec:birefringence}

The fact that in the ``LIV extended QED" discussed in section \ref{sec:MP} opposite ``helicities'' have slightly different group velocities, implies that the polarisation vector of a linearly polarised plane wave with energy $k$ rotates, during the wave propagation over a distance $d$, through the angle 
\begin{equation} 
\theta(d) = \frac{\omega_{+}(k)-\omega_{-}(k)}{2}d \simeq \xi^{(3)}\frac{k^2 d}{2\,M_{\rm Pl}}\;.
\label{eq:theta}
\end{equation} 
It is obvious that such angle will be different for different photon energies hence this effect could potentially disrupt the amount of polarisation present in a some polarised light travelling over long distances. More specifically, depending on the amount of available information on the specific object observed, one can use this fact to cast constraints on the photon LIV parameter $|\xi^{(3)}|$ in two different ways \cite{Maccione:2008tq}:

\paragraph{Decrease in polarization degree.}
If some net amount of polarization is measured in the band say $k_{1} < E < k_{2}$, an order-of-magnitude constraint arises from the fact that the angle of polarization rotation (\ref{eq:theta}) cannot differ by more than $\pi/2$ over this band, as in this case the detected polarization would fluctuate sufficiently for the net signal polarization to be suppressed \cite{Gleiser:2001rm, Jacobson:2003bn}.   From (\ref{eq:theta}), assuming $\Delta\theta \leq \pi/2$ implies
\begin{equation} 
\xi^{(3)}\lesssim\frac{\pi\,M_{\rm Pl}}{(k_2^2-k_1^2)d(z)}\;,
\label{eq:decrease_pol}
\end{equation} 
This constraint merely relies on the detection of a polarized signal and does not make use of the observed
polarization degree but one can obtain for a more refined limit by calculating the maximum
observable polarization degree, given some maximum intrinsic value \cite{McMaster}:
\begin{equation} 
\Pi(\xi) = \Pi(0) \sqrt{\langle\cos(2\theta)\rangle_{\mathcal{P}}^{2}
+\langle\sin(2\theta)\rangle_{\mathcal{P}}^{2}},
\label{eq:pol}
\end{equation} 
where $\Pi(0)$ is the maximum initial degree of polarization,
$\theta$ is defined in Eq.~(\ref{eq:theta}) and the average is
weighted over the source spectrum and instrumental efficiency,
represented by the normalized weight function
$\mathcal{P}(k)$~\cite{Gleiser:2001rm}.  
Conservatively, one can a priori set $\Pi(0)=100\%$, however for many astrophysical sources a lower value 
may be justified on the basis of theoretical modelling.
Using (\ref{eq:pol}), one can then 
cast a constraint by 
requiring $\Pi(\xi)$ to exceed the observed value. 

\paragraph{Rotation of polarization angle.}
A different possibility for casting a constraint consists in looking at the overall rotation of the average polarisation.
Let's suppose that one observes polarised light over some energy band which shows an overall position angle $\theta_{\rm obs}$ with respect to a fixed
direction. At fixed energy, the polarization vector rotates by the
angle (\ref{eq:theta}) (Faraday rotation is negligible at
high energies.); if the position angle is measured by averaging over a
certain energy range, the final net rotation 
$\left<\Delta\theta\right>$
is given by the superposition of the polarization vectors of all the photons in that
range:
%
%
\begin{equation}
\tan (2\left\langle\Delta\theta\right\rangle) = \frac{
\left\langle\sin(2\theta)\right\rangle_{\mathcal{P}}}{\left\langle
\cos(2\theta)\right\rangle_{\mathcal{P}}}\;,
\label{eq:caseB}
\end{equation}
where 
$\theta$ is given by (\ref{eq:theta}).
If the position angle at emission $\theta_{\rm i}$ in the same energy band 
is known from a model of the emitting source, a constraint can be set by imposing
\begin{equation}
\tan(2\left\langle\Delta\theta\right\rangle) < \tan(2\theta_{\rm obs}-2\theta_{\rm i})\;.
\label{eq:constraint-caseB}
\end{equation}
%
While this limit is tighter than those based on eqs.~(\ref{eq:decrease_pol}) and (\ref{eq:pol}), it clearly hinges on assumptions about the 
nature of the source, 
which may introduce significant uncertainties.

In conclusion the fact that polarised photon beams are indeed observed from distant objects imposes strong constraints on LIV in the photon sector (i.e.~on $\xi^{(3)}$), as we shall see later on.

\subsubsection{Threshold reactions\index{Threshold reactions}}
\label{sec:thresholds}

LIV corrections are quite important in threshold processes because the LIV term (which as a first approximation can be considered as an additional mass term) only needs to be comparable to the (invariant) mass of the particles produced in the final state for strongly affecting this kind of reactions. 
I.e. a rough estimate for the threshold energy is
\begin{equation}
p_{\rm th} \simeq \left(\frac{m^{2}\Mpl^{n-2}}{\eta^{(n)}}\right)^{1/n}\;,
\label{eq:threshold-general}
\end{equation}
where $m$ is the typical mass of particles involved in the reaction. Interesting values for $p_{\rm th}$ are discussed, e.g., in \cite{Jacobson:2002hd} and given in Tab.~\ref{tab:thresholds}. 
%
\begin{table}[htbp]
\caption{Values of $p_{\rm th}$, according to eq.~(\ref{eq:threshold-general}), for different particles involved in the reaction: neutrinos, electrons and proton. Here we assume $\eta^{(n)} \simeq 1$.}
\begin{center}
\begin{tabular}{|c|c|c|c|}
\hline
& $m_{\nu}\simeq 0.1$~eV & $m_{e}\simeq 0.5$~MeV & $m_{p} \simeq 1$~GeV \\
\hline
$n=2$ & 0.1 eV & 0.5 MeV & 1 GeV\\
\hline 
$n=3$ & 500 MeV & 14 TeV & 2 PeV\\
\hline 
$n=4$ & 33 TeV & 74 PeV & 3 EeV\\
\hline
\end{tabular}
\end{center}
\label{tab:thresholds}
\end{table}%
Reactions involving neutrinos are the best candidate for observation of LIV effects, whereas electrons and positrons can provide results for $n=3$ theories but can hardly be accelerated by astrophysical objects up to the required energy for $n=4$. In this case reactions of protons can be very effective, because cosmic-rays can have energies well above 3 EeV.

For the above reason a quite rich phenomenology of threshold reactions is introduced by LIV in EFT and threshold theorems can be generalized \cite{Mattingly:2002ba}. Sticking to the present case of rotational invariance and monotonic dispersion relations (see \cite{Baccetti:2011us} for a generalization to more complex situations), the main conclusions of the investigation into threshold reactions are that \cite{Jacobson:2002hd}
\begin{itemize}
\item Threshold configurations still corresponds to head-on incoming particles and parallel outgoing ones
\item The threshold energy of existing threshold reactions can shift, and upper thresholds (i.e.~maximal incoming momenta at which the reaction can happen in any configuration) can appear
\item Pair production can occur with unequal outgoing momenta
\item New, normally forbidden reactions can be viable
\end{itemize}

Let us now briefly review the main reaction used so far in order to casts constraints.

\subsubsection{LIV-allowed threshold reactions: $\gamma$-decay\index{$\gamma$-decay}} 
While normally the decay of a photon in a lepton-antilepton pair is kinematically forbidden, it might be allowed in the presence of LIV induce modified dispersion relations of the sort we considered so far.  Taking the case $n=3$ and $\xi^{(3)}=0$ as a simple example, the threshold is set by the condition \cite{Jacobson:2005bg}
\begin{equation}
k_{th} = (6\sqrt{3}m_e^2M_{\rm Pl}/|\eta_\pm^{(3)}|)^{1/3}\, . 
\end{equation}
Noticeably, as already mentioned above, the electron-positron pair can now be created with slightly different outgoing momenta (asymmetric pair production)~\cite{Jacobson:2002hd,Mattingly:2002ba}. Furthermore, the decay rate is extremely fast above threshold \cite{Jacobson:2005bg} and is of the order of $(10~{\rm ns})^{-1}$ ($n=3$) or $(10^{-6}~{\rm ns})^{-1}$ ($n=4$).

\subsubsection{LIV-allowed threshold reactions: Vacuum \v{C}erenkov\index{Vacuum \v{C}erenkov} and Helicity Decay\index{Helicity Decay}} 
In the presence of LIV, also the process of Vacuum \v{C}erenkov (VC) radiation $e^{\pm}\rightarrow e^{\pm}\gamma$ can occur (this can be easily seen, as for the gamma decay, by imposing the energy momentum conservation). If we set again $n=3$ and $\xi^{(3)} \simeq 0$ and , the threshold energy is given by
\begin{equation}
 p_{\rm VC} = (m_e^2M_{\rm Pl}/2\eta^{(3)})^{1/3} \simeq 11~\mbox{TeV}~\eta^{-1/3}\;.
\label{eq:VC_th}
\end{equation}
Just above threshold this process is also extremely efficient, with a time scale of order $\tau_{\rm VC} \sim 10^{-9}$~s \cite{Jacobson:2005bg}. 

A slightly different version of this process is the Helicity Decay (HD, $e_{\mp}\rightarrow e_{\pm}\gamma$). If $\eta_{+} \neq \eta_{-}$, an electron/positron can flip its helicity by emitting a suitably polarized photon. This reaction does not have a real threshold, but rather an effective one \cite{Jacobson:2005bg} --- $ p_{\rm HD} = (m_e^2M_{\rm Pl}/\Delta\eta)^{1/3}$, where $\Delta\eta = |\eta_+^{(3)}-\eta_-^{(3)}|$ --- at which the decay lifetime $\tau_{HD}$ is minimized. For $\Delta\eta\approx O(1)$ this effective threshold is around 10 TeV.  
Note that below threshold $\tau_{\rm HD} > \Delta\eta^{-3} (p/10~\mbox{TeV})^{-8}\, 10^{-9}$s, while above threshold $\tau_{\rm HD}$ becomes independent of $\Delta\eta$~\cite{Jacobson:2005bg}. 

Apart from the above mentioned examples, one can also look for modifications of normally allowed threshold reactions especially relevant in high energy astrophysics.

\subsubsection{LIV-allowed threshold reactions: photon splitting and lepton pair production}
\label{subsec:ph-spl-el-pair}

Once photon decay and vacuum \v{C}erenkov are allowed also the related relations in which respectively the our going lepton pair is replaced by two or more photons, $\gamma \rightarrow 2 \gamma$ and $\gamma \rightarrow 3 \gamma$, etc.\ , or the outgoing photons is replaced by an electron-positron pair, $e^- \rightarrow e^-e^-e^+$, are also allowed.

\paragraph{Photon splitting\index{Photon Splitting}}

It is easy to show that this process requires UV superluminal photons, i.e. $\xi^{(n)} > 0$ \cite{Jacobson:2002hd}. When allowed, the relevance of this process is simply related to its rate. The most relevant cases are $\gamma\rightarrow \gamma\gamma$ and $\gamma\rightarrow3\gamma$, because processes with more photons in the final state would be suppressed by more powers of the fine structure constant. 

The $\gamma\rightarrow\gamma\gamma$ process is forbidden in standard QED because of kinematics and CP conservation. In LIV QED neither condition holds, however, we can argue that this process is suppressed by an additional power of the Planck mass, with respect to $\gamma\rightarrow 3 \gamma$. In fact, in LI QED the matrix element is zero due to the exact cancellation of fermionic and anti-fermionic loops. In LIV EFT this cancellation is not exact and the matrix element is expected to be proportional to at least $\xi (E/\Mpl)^{p}$, $p>0$, as it is induced by LIV and must vanish in the limit $\Mpl \rightarrow \infty$. 

Let us consider then $\gamma\rightarrow 3 \gamma$. This process has been studied for $n=3$ in \cite{Jacobson:2002hd,Gelmini:2005gy}. In particular, in \cite{Gelmini:2005gy} it was found that, if the ``effective photon mass'' $m_{\gamma}^{2} \equiv \xi E_{\gamma}^{n}/\Mpl^{n-2} \ll m_{e}^{2}$, then the splitting lifetime of a photon is approximately $\tau^{n=3}\simeq 0.025\,\xi^{-5} f^{-1}\left(50~\mbox{TeV}/E_{\gamma}\right)^{14}$~s, where $f$ is a phase space factor of order 1.  This rate was rather higher than the one obtained via dimensional analysis in \cite{Jacobson:2002hd} because, due to integration of loop factors, additional dimensionless contributions proportional to $m_{e}^{8}$ enhance the splitting rate at low energy.

\paragraph{Lepton pair production}
The process $e^- \rightarrow e^-e^-e^+$ is similar to vacuum
\v{C}erenkov radiation or helicity decay, with the final photon
replaced by an electron-positron pair.   While various combinations of helicities for the different fermions can be considered, we shall consider here, for illustrative reasons, the most simple case where all electrons have the same helicity and the positron has the opposite helicity, so that the threshold energy will depend on only one LIV parameter. The threshold for this reaction was derived in  In \cite{Jacobson:2002hd} and found higher than that for soft vacuum \v{C}erenkov radiation by a factor $2.5$. Given that the reaction rate is high as well, one can casts constraints by using just the value of the threshold.

\subsubsection{LIV-modified threshold reactions: Photon pair-creation\index{Pair-creation}}
\label{sec:ph-pair}

A process related to photon decay is photon absorption, $\gamma\gamma_0\rightarrow e^+e^-$. Unlike photon decay,
this is allowed in Lorentz invariant QED and it plays a crucial role in making our universe opaque to gamma rays above tents of TeVs as these are absorbed in this way by the infra-red and cosmic microwave background radiation. 

Let's call $\omega_0$ the (low) energy associated to the target photon $\gamma_0$, then the threshold for the reaction occurs in a
head-on collision with the second photon having momentum $k_{\rm LI}=m^2/\omega_{0}$. 
For example, if $k_{\rm LI}=10$ TeV (the typical energy of inverse Compton generated photons in some active galactic nuclei) the
soft photon threshold $\omega_0$ is approximately 25 meV, corresponding to a wavelength of 50 microns.

In the presence of Lorentz violating dispersion relations the threshold for this process is in general altered, and the process
can even be forbidden. Moreover, as we mentioned before, in some cases an upper threshold con develop for which the process does not occur~\cite{Kluzniak:1999qq,Jacobson:2002hd,Mattingly:2002ba,Baccetti:2011us}. Physically, this means that at sufficiently high momentum the photon does not carry enough energy to create a pair and simultaneously conserve energy and momentum. Note also, that an upper threshold can only be found in regions of the parameter space in which the $\gamma$-decay is forbidden, because if a single photon is able to create a pair, then {\em a fortiori} two interacting photons will do \cite{Jacobson:2002hd}. 

Let us exploit the above mentioned relation $\eta_{\pm}^{e^{-}} = (-)^{n}\eta_{\mp}^{e^{+}}$ between the electron-positron coefficients, and assume that on average the initial state is unpolarized. In this case, using the energy-momentum conservation, the kinematics equation governing pair production is the following \cite{Jacobson:2005bg}
\begin{equation}
\frac{m^{2}}{k^{n}y \left(1-y\right)} =  
  \frac{4\omega_{0}}{k^{n-1}} + {{\xi^{(n)}}\over{M_{\rm Pl}^{n-2}}} - {{\eta^{(n)}}\over{M_{\rm Pl}^{n-2}} }\left( y^{n-1}+(-)^{n}\left(1-y\right)^{n-1}\right)
   \label{eq:ggscat}
\end{equation}
where $0 < y < 1$ is the fraction of momentum carried by either the electron or the positron with respect to the momentum $k$ of the incoming high-energy photon. In general the  analysis is rather complicated. In particular it is necessary to sort out whether the thresholds are lower or upper ones, and whether they occur with the same or different pair momenta

\subsubsection{Synchrotron radiation\index{Synchrotron radiation}} 
\label{synchrotron}
Synchrotron emission is strongly affected by LIV~\cite{Jacobson:2002ye,Ellis:2003sd,Montemayor:2004mh,Montemayor:2005ka,Jacobson:2005bg,Altschul:2006pv,Maccione:2007yc,Liberati:2012jf}, however for Planck scale LIV and observed energies, it is a relevant ``window of opportunity" for QG phenomenology only for dimension four or five LIV QED albeit we shall see that in specific QG models (see below section \ref{sec:horava}), with the Lorentz breaking scale not equal {\em a priori} to the Planck one, it might play a relevant role also for dimension six LIV operators. For this reason we shall generically indicate the Lorentz breaking scale $M_{LIV}$ in what follows.

In both LI and LIV cases \cite{Jacobson:2005bg}, most of the radiation from an electron of energy $E$ is emitted at a critical frequency
\begin{equation}
 \omega_c = \frac{3}{2}eB\frac{\gamma^3(E)}{E} 
\label{eq:omega_sync}
\end{equation}
where $\gamma(E) = (1-v^2(E))^{-1/2}$, and $v(E)$ is the electron
group velocity. Assuming Hamiltonian dynamics, the group velocity of an electron with a modified dispersion relation at arbitrary order as in (\ref{eq:mdrqed}) can be computed as (neglecting higher order terms)~\cite{Liberati:2012jf}
\begin{equation}
v(E) = \frac{\partial E}{\partial p} \simeq 1-\frac{m^{2}}{2p^{2}} + \frac{n-1}{2}\eta\left(\frac{p}{M_{\rm LIV}}\right)^{n-2}.
\end{equation}
We immediately see that $v(E)$ can exceed $1$ if $\eta > 0$ or it can be strictly less than $1$ if $\eta < 0$ once the LIV term is prevailing on the suppressed mass term. This introduces a fundamental difference between particles with positive or negative LIV coefficient $\eta$ and the possibility to see the effect even for $p/M_{\rm LIV}\ll 1$.  In \cite{Jacobson:2002ye} a simple, but stringent, constraint for $n=3$ was cast using the fact that subluminal dispersion relations ($\eta<0$) admit a maximal critical frequency for the synchrotron emission. For arbitrary $n>2$ this would be~\cite{Liberati:2012jf}
\begin{equation}
\omega_{c}^{max,(n)}=\frac{3eB}{2m}\left(1-\frac{4}{3n}\right)^{3/2}\!\!\left(\frac{4\, (M_{\rm LIV}/m)^{n-2}}{-\eta(n-1)(3n-4)}\right)^{2/n}.
\label{eq:ommax}
\end{equation}
Basically, if synchrotron emission up to some frequency $\omega_{\rm obs}$ is observed, one can deduce that the LIV coefficient for the corresponding leptons cannot be more negative than the value for which $\omega_c^{\rm max}=\omega_{\rm obs}$. 

If $\eta$ is instead positive the leptons can be superluminal. One can show, e.g. for $n=3$ and $M_{\rm LIV}=M_{\rm PL}$, that at energies $E_c \gtrsim 8~\mbox{TeV} /\eta^{1/3}$, $\gamma(E)$ begins to increase faster than $E/m_e$ and reaches infinity at a finite energy, which corresponds to the threshold for soft VC emission. The critical frequency is thus larger than the LI one and the spectrum shows a characteristic bump due to the enhanced $\omega_c$. 
While this is the general behaviour to be expected, a derivation of a constraint in this case requires a direct confrontation with observation and hence detailed reconstruction of the synchrotron spectrum in the presence of LIV effects~\cite{Maccione:2007yc,Liberati:2012jf}.

\section{Experimental and observational constraints}

We shall now give a brief overview of the most stringent constraints available on the different sector of the SME. Of course, different operators, of different mass dimension, require different sort of observations and hence it will make sense to divide our discussion not only by SM sector but also by the order of the operators considered. 

\subsection{Rotational invariant QED in mSME}
Lorentz and CPT breaking operators does not depend on energy in the mSME.  Therefore it is generally true that  high precision laboratory and terrestrial experiments might be preferable with respect to astrophysical observation in this case.  For a recent listing of all the constraints including the non-rotationally invariant case, see~\cite{Kostelecky:2008ts}, we shall limit ourself here to the rotationally invariant case discussed before.

\subsubsection{Spin polarized torsion constraints on electrons:}
Spin polarized torsion balances can place limits on the electron sector of the mSME~\cite{Heckel:2008hw} and can be very roughly described as a magnet attached on a torsion fiber. Hence, torsion balances are characterised by a large number of aligned electron spins. The mSME coefficients give rise to an interaction potential for non-relativistic electrons which produces an orientation dependent torque on the torsion balance which can be measured using the twist of the torsion fiber.  The torsion balance in its sealed vacuum chamber is mounted on a rotating turntable, which allows for very sensitive detection of any anomalous torque as a function of the rotation frequency (and the Earth's rotation and motion in the solar system). These experiments sets limits on the $b$ coefficient in the electron modified Lagrangian Eq.~(\ref{eq:LIVQEDelectronrotinv}). Current limits are of the order of $10^{-27}$ GeV~\cite{Heckel:2008hw}.

\subsubsection{Accelerator constraints on electrons:}
Energy loss mechanisms such as the vacuum \v{C}erenkov effect $q \rightarrow q + \gamma$ (see section \ref{sec:thresholds}) for charged fermions are incompatible with a constant beam energy as long as the energy loss rate is high enough. The LEP experiment accelerated electrons and positrons to energies of roughly 100 GeV in the lab frame and it is known that the beam does not lose any significant energy to vacuum \v{C}erenkov radiation.  Additionally the synchrotron emission from the beam has been measured.  Using the characteristics of the LEP beam and the related synchrotron radiation spectrum sets limits on the coefficient $c$ for electrons (as in Eq.~(\ref{eq:LIVQEDelectronrotinv})) to be of order $|c|<10^{-15}$.

\subsubsection{Astrophysical inverse Compton bounds on electrons:}
The parameter $d$ in Eq.~(\ref{eq:LIVQEDelectronrotinv}) controls  a spin dependent change to the dispersion relation for a fermion. We observe energies from the radio up to 80 TeV from severe active galactic nuclei as well as pulsar wind nebula, e.g. the Crab nebula~\cite{Aharonian:2004gb}.  The overall spectrum of these sources is well understood and the high energy emission is dominated by inverse Compton scattering of accelerated electrons off of lower energy photons (mainly synchrotron and CMB photons depending on the specific astrophysical object).  The magnetic field present in the same sources causes the electrons to precess, thereby destroying any initial polarization of the electrons.  Therefore, one can argue that both helicities of electrons must be present and stable if there is to be inverse Compton radiation~\cite{Altschul:2006he}.  This fact forbids the vacuum \v{C}erenkov effect for both helicities, implying a double sided constraints on $d$ of the order of $|d|<O(10^{-12})$~\cite{Altschul:2006he} in the CMB rest frame.

\subsubsection{Cosmic ray and HESS bounds on photons:}
The observation of high energy cosmic ray protons and TeV gamma rays allows to cast constraints on the  on $k_F$ parameter for photons in Eq.~(\ref{eq:LIVQEDphotonrotinv}). Assuming unmodified protons (within the mSME which sector is the ``unmodified'' sector is arbitrary, in that the limiting speed of one of the sectors can be defined to be the ``speed of light''), in~\cite{Klinkhamer:2008ky} the authors used the necessary absence of vacuum \v{C}erenkov $p\rightarrow p + \gamma$ and a Pierre Auger event in conjunction with the excess of high energy TeV gamma rays observed by the HESS telescope (which forbids the gamma decay $\gamma \rightarrow p+\overline{p}$) to produce a two sided bound on $k_F$ of $-9 \times 10^{-16} < k_F< 6 \times 10^{-20}$.

\subsubsection{Colemann-Glashow limit of the mSME QED:}
One can of course restrict the QED sector in the mSME to the rotational invariant subset and solely dimension four operators. In this case the model basically coincides with the Coleman-Glashow one~\cite{Coleman:1998ti}. Also in this case the constraints are quite strong, for example on the QED sector one can easily see that the absence of gamma decay up to 50 TeV provides a constraint of order $10^{-16}$ on the difference between the limit speed of photons and electrons~\cite{Stecker:2001vb}. Similarly, the observations of protons up to $10^{19}$~eV in cosmic rays is sufficient evidence for a constraint of order $10^{-20}$ on the difference of speed between them and light (constraints of order $10^{-22}\div10^{-23}$ can also be achieved, and are sometimes quoted in the literature, if one takes slightly higher values for the observed energies of propagating protons. We report here a more conservative constraint given the present uncertainties on the highest energy tail of the cosmic rays spectrum, see Section~\ref{open-issue}).

\subsection{QED with rotational invariant CPT odd dimension 5 ops}

There is a very rich literature devoted to constraining the Myers-Pospelov version of the SME. We shall not attempt here a detailed discussion but just briefly summarise the available constraints.

\subsubsection{Synchrotron radiation constraints from the Crab Nebula:}
Presently a very important object in this sense is the Crab nebula (CN). This pulsar wind nebula had its origin by a supernova explosion observed in 1054~A.D. Its distance from Earth is approximately $1.9$~kpc and it is one of the best studied sources of diffuse radio, optical and X-ray radiation. The Nebula emits an extremely broad-band spectrum (21 decades in frequency, see \cite{Maccione:2007yc} for a comprehensive list of relevant observations) that is produced by two major radiation mechanisms.  The emission from radio to low energy $\gamma$-rays ($E < 1$~GeV) is thought to be synchrotron radiation from relativistic electrons,  whereas inverse Compton (IC) scattering by these electrons is the favoured explanation for the higher energy  $\gamma$-rays.   From a theoretical point of view, the current understanding of the whole environment is based on the so called Kennel--Coroniti~\cite{Kennel:1984vf}, which accounts quite accurately for the general features observed in the CN spectrum. 

Developing on the initial idea discussed in \cite{Jacobson:2002ye}, a full reconstruction of the synchrotron emission processes in the CN within the SME with dimension 5 CPT operators (see Eqs.~(\ref{eq:disp_rel_phot}) and (\ref{eq:disp_rel_ferm})) has been performed in \cite{Maccione:2007yc}. 
This procedure requires fixing most of the model parameters using radio to soft X-rays observations (which are basically unaffected by LIV), a careful step by step revisitation of the basic assumptions made in the Kennel--Coroniti model and also to take into account the contribution of novel LIV effects such as  vacuum \v{C}erenkov and helicity decay. Finally, a $\chi^2$ analysis can be performed to quantify the agreement between models and data \cite{Maccione:2007yc}. From this analysis, one can conclude that the LIV parameters for the leptons are both constrained, at 95\% CL, to be $|\eta_\pm| < 10^{-5}$.  

\subsubsection{Birefringence constraints:}
The best constraints on the photon sector are instead obtained by using birefringence effects associated with the CPT odd nature of the relevant LIV operators. Strong constraints came again from the CN \cite{Maccione:2008tq}, from which a value $|\xi^{(3)}| \lesssim 6 \times 10^{-10}$ at 95\% Confidence Level (CL) was obtained by considering the observed polarization of hard-X rays \cite{integralpol} (see also \cite{Forot:2008ud}).
Polarized light from GRBs has also been detected and given their cosmological distribution they could be ideal sources for improving the above mentioned constraints from birefringence. Attempts in this sense were done in the past \cite{Jacobson:2003bn,Mitro} but the observed of polarisation was later deemed controversial. Furthermore, GRBs for which the polarization is detected and the spectral redshift is precisely determined are scarce. In \cite{Stecker:2011ps} this problem was circumvented by using indirect methods (the same used to use GRBs as standard candles) for the estimate of the redshift. This leads to a possibly less robust but striking constraint $|\xi^{(3)}| \lesssim 2.4\times10^{-15}$.  
A similar constraint $|\xi^{(3)}| \lesssim 10^{-15}$ was obtained in~\cite{Toma:2012xa} by making use of two GRBs observed by the Japanese satellite IKAROS. Also in this case the redshift was determined using indirect methods such as the correlation between peak energy and luminosity of the GRB prompt emission. Remarkably, the above mentioned constraints were recently improved by using the INTEGRAL/IBIS observation of the GRB 061122, for which a redshift $z\approx 1.33$ was derived thanks to the determination of the GRB's host galaxy. In this case a constraint $|\xi^{(3)}| \lesssim 3.4\times10^{-16}$ was derived \cite{Gotz:2013dwa} (see also~\cite{Laurent:2011he} for a similar previous constraint).

\subsection{QED with rotational invariant CPT even dimension 5 and 6 ops}
\label{sec:n4constr}

As we have shown before, this kind of operators lead to modified dispersion relations characterised by quartic terms in the particle momenta suppressed by the squared Planck mass. Furthermore the CPT even nature of the operators prevent birefringent features in the photon sector. This feature basically implies the impossibility to use any of the above discussed observation for casting effective constraints (e.g. the synchrotron emission from the CN would not provide a constraint better that $10^6$ on the electron parameter given the $(E/M_{\rm Pl})^2$ suppression). However, there are available observations exploring much higher energies, i.e. those related to the so called Ultra High Energy Cosmic Rays (UHECR). 

\subsubsection{Astrophysical constraints from GZK reaction secondaries:}
The GZK cut off \cite{Greisen:1966jv,1969cora...11...45Z}, is a suppression of the high-energy tail of the UHECR spectrum arising from interactions with CMB photons: $p\gamma\rightarrow \Delta^{+}\rightarrow p\pi^{0}(n\pi^{+})$. When Lorentz invariance holds, this process has a threshold energy $E_{\rm th} \simeq 5\times 10^{19}~(\omega_{0}/1.3~\mbox{meV})^{-1}$~eV (where $\omega_{0}$ is the target photon energy). Experimentally, the presence of a suppression of the UHECR flux was claimed only recently \cite{Abbasi:2007sv,Roth:2007in} (and, as discussed in section~\ref{sec:hist}, it was initially claimed to be absent by the AGASA collaboration~\cite{Takeda:1998ps}). Although the cut off could be also due to the finite acceleration power of the UHECR sources, the fact that it occurs at the expected energy favours the GZK explanation. The results presented in \cite{Cronin:2007zz} seemed to further strengthen this hypothesis (but see further discussion in the conclusions).

Significant limits on $\xi=\xi^{(4)}$ and $\eta=\eta^{(4)}$ for the electron/positron can be derived by considering UHE photons generated as secondary products of the GZK reaction\cite{Galaverni:2007tq,Maccione:2008iw}. These UHE photons originate because the GZK process leads to the production of neutral pions that subsequently decay into photon pairs. These photons are mainly absorbed by pair production onto the CMB and radio background. Thus, the fraction of UHE photons in UHECRs is theoretically predicted to be less than 1\% at $10^{19}$~eV \cite{Gelmini:2005wu}. Several experiments imposed limits on the presence of photons in the UHECR spectrum. In particular, the photon fraction is less than 2.0\%, 5.1\%, 31\% and 36\% (95\% C.L)~at $E = 10$, 20, 40, 100 EeV  respectively \cite{Aglietta:2007yx,Rubtsov:2006tt} (at lower energies even stricter limits 0.4\%, 0.5\%, 1.0\%, 2.6\% and 8.9\% (95\% C.L) above $E=1$, 2, 3, 5 and 10 EeV respectively were recently found \cite{Kampert:2012vh}).

The key idea for casting constraints here, relies on the strong dependence of the photon pair production on on LIV modifications (see section~\ref{sec:ph-pair}). In particular, the (lower) threshold energy can be shifted and in general an upper threshold can be introduced \cite{Jacobson:2002hd}. If the upper threshold energy is lower than $10^{20}$~eV, then GZK secondary UHE photons are no longer attenuated by the CMB and can reach the Earth. Hence, they would constitute a significant fraction of the total UHECR flux and thereby they would  violate the above mentioned experimental bounds \cite{Galaverni:2007tq,Maccione:2008iw,Galaverni:2008yj}.  Of course, one LIV is introduced these extra UHE photons could now be removed via the above discussed photon splitting process. However, the analysis of the rate we presented in \cite{Gelmini:2005gy} and discussed in section~\ref{subsec:ph-spl-el-pair}, does not apply to photons around $10^{19}$ eV given that at these energies $m_{\gamma}^{2} \gg m_{e}^{2}$ if $\xi^{(3)} > 10^{-17}$ and $\xi^{(4)} > 10^{-8}$. In this regime, contributions to the rate are in this case at most logarithmic, as the momentum circulating in the fermionic loop is much larger than $m_{e}$. Moreover, the splitting rate depends only on $m_{\gamma}$, the only energy scale present in the problem. One then expects the analysis proposed in \cite{Jacobson:2002hd} to be correct and the splitting time scale to be negligible at $E_{\gamma} \simeq 10^{19}$~eV.

To complete the analysis one can also notice that the $\gamma$-decay process can also imply a significant constraint. Indeed, if some UHE photon ($E_{\gamma}\simeq 10^{19}$~eV) is detected by experiments (and the Pierre Auger Observatory, PAO, will be able to do so in few years \cite{Aglietta:2007yx}), then $\gamma$-decay must be forbidden above $10^{19}$~eV~ \cite{Maccione:2008iw} (note that for $n=4$ the two photons felicities travel at the same speed, so there is in this case a single population).


The combination of above mentioned ``upper threshold constraint" and the expected constraint from the absence of gamma decay would bound the QED LIV parameters at order $n=4$ to the roughly rectangular region (see Figure~2, of~\cite{Liberati:2009pf} for more details)
\begin{eqnarray}
-10^{-7} \lesssim &\xi^{(4)} \lesssim 10^{-8}\qquad-10^{-7} \lesssim &\eta^{(4)} \lesssim 10^{-6}
\end{eqnarray}

\subsection{SME: Hadronic sector}

The very same GZK photo-pion production is strongly affected by LIV and the constraints implied by the detection of  this effect have been extensively considered in the literature~\cite{Aloisio:2000cm,Alfaro:2002ya,Jacobson:2002hd,Mattingly:2008pw,Scully:2008jp,Stecker:2009hj}. Nonetheless, a detailed LIV study of the GZK feature is hard to perform, because of the many astrophysical uncertainties related to the modelling of the propagation and the interactions of UHECRs. In fact, the Lorentz breaking operators could lead to relevant modifications of the GZK mean free path. Consequently, the propagated UHECR spectrum can display new features, like bumps at specific energies, suppression at low energy, recovery at energies above the cutoff. These are all features which cannot be easily conciliated with the observed spectrum, even taking into account experimental uncertainties. Furthermore, the emission of Cherenkov $\gamma$-rays and pions in vacuum would lead to sharp suppression of the spectrum above the relevant threshold energy. After a detailed statistical analysis of the agreement between the observed UHECR spectrum and the theoretically predicted one in the presence of LIV and assuming pure proton composition, the final constraints implied by UHECR physics are (at 99\% CL) \cite{Maccione:2009ju}
\begin{eqnarray}
\nonumber
-10^{-3} \lesssim &\eta^{(4)}_{p}& \lesssim 10^{-6}\\
-10^{-3} \lesssim &\eta^{(4)}_{\pi}&  \lesssim 10^{-1} \quad (\eta^{(4)}_{p} > 0) \quad\mbox{or}\quad \lesssim 10^{-6} \quad (\eta^{(4)}_{p} < 0)\,.
\label{eq:finalconstraint}
\end{eqnarray}
Obviously the same analysis can be applied also to dimension five, CPT odd, operators leading to much stronger constraints (order $O(10^{-14})$).

\subsection{An open issue from UHECR}
\label{open-issue}

We just saw how the GZK cutoff in UHECR physics has played a determinant role in casting constraints for $n=4$ (CPT even dimension 5 and 6 operators) dispersion relations in the QED and Hadronic sectors of the SME. It is hence quite unfortunate that these constraints are still suffering from experimental uncertainties about the composition of the UHECR and hence the nature of the steep fall-off in the spectrum observed above $10^{19}$  eV. Indeed, UHECR constraints have relied so far on the hypothesis that protons constituted the majority of UHECRs above $10^{19}$~eV so that the step fall off detected at high energies could be univocally identified with the GZK one. 

Recent AUGER \cite{Abraham:2010yv,Kampert:2012vh} and Yakutsk \cite{Glushkov:2007gd} observations, however, showed hints of an increase of the average mass composition with rising energies up to $E \approx 10^{19.6}$~eV, although still with large uncertainties mainly due to the proton-air cross-section at ultra high energies. Hence, experimental data suggests that heavy nuclei can possibly account for a substantial fraction of UHECR arriving on Earth. This would be in agreement with fact that the initial evidence~for correlations between UEHCR events and active galactic nuclei (AGN) --- mainly Blasars ---  observed by AUGER~\cite{Cronin:2007zz} has not improved (actually has decreased) with increasing statistics~\cite{Kampert:2012vh}. 

Indeed, heavy ions are much more deviated by the extra and inter galactic magnetic fields due to their larger charge with respect to protons (albeit this effect is partially compensated by their shorter mean free path at very high energies). The situation is made even more confused by the fact that these AUGER observations seem at odd with those of the Telescope Array (TA) based in Utah (hence in the northern hemisphere, while AUGER is based in Argentina) which indicate a UHECR composition dominated by protons~\cite{1742-6596-375-5-052007} and some hint of correlation (the TA will need to reach about 100 events to detect with statistic significance, at the moment about 40 vents are available) with large scale structures in the universe for UHECR with $E>57$ EeV (albeit also TA does not see any significant correlation with AGN)~\cite{AbuZayyad:2012hv}.

Given the above described uncertainties, it is worth considering how our constraints on dimension 5 and 6 CPT even operators would change if the UHECR composition would be in the end be dominated by heavy ions. The first bad news would be the loss of the constraints at $n=4$ on the QED sector as they are based on the proton GZK reaction. Actually, to be fair, this is perhaps a too harsh statement. In fact, it was shown in \cite{Hooper:2010ze} that at some high energy gamma ray flux is still expected also in the case of mixed composition (albeit reduced with respect to the purely proton one). Hence, the previously discussed line of reasoning based on the absence of upper threshold for UHE gamma rays might still work. 

For what regards the hadronic sector, UHE nuclei suffer mainly from photo-disintegration losses as they propagate in the intergalactic medium. Of course, also photo-disintegration is a threshold process and can be strongly affected by LIV. According to \cite{Saveliev:2011vw}, the mean free paths of UHE nuclei are modified by LIV in such a way that the final UHECR spectra can show distinctive LIV features. However, a quantitative evaluation of the propagated spectra has not been performed yet.
An interesting constraint was nonetheless derived in~\cite{Saveliev:2011vw} assuming, for simplicity, a single Lorentz invariance violation parameter $\bar{\eta}^{(4)}$ controlling Lorentz invariance violation for all nuclear species. A lower limit on $\bar{\eta}^{(4)}$ was set by requiring that nuclei, which are stable in the Lorentz invariant case, should not undergo spontaneous decay (by emitting single nucleons) below energies of the order of $10^{19.5}$ to $10^{20}$ eV, given that at these energies a substantial fraction of heavy ultra high energy cosmic ray nuclei has been observed. Conversely, an upper limit was given by requiring the absence of vacuum Cherenkov emission at the same energies. The final constraints in this case were found to be $-3\times10^{-2} \lesssim \bar{\eta}^{(4)} \lesssim 4$ for $^{56}$Fe, $-2\times 10^{-3} \lesssim \bar{\eta}^{(4)} \lesssim 3 \times 10^{-2}$ for $^{16}$O and $-7 \times 10^{-5} \lesssim \bar{\eta}^{(4)} \lesssim 1 \times 10^{-4}$ for $^4$He, respectively.

A final comment is deserved for the UHECR atmospheric showers reconstruction techniques. While preliminary, incomplete studies, hinted toward some robustness agains LIV modifications of the overall shower observables, there is a growing activity aiming at a more precise reanalysis of the problem~\cite{Rubtsov:2012kb,Satunin:2013an}. This is crucial to understand given that, for example, the absence of observation of UHE photons could be mimicked by anomalous shower developments which would then fool us about the nature of the impinging UHE particles. New results are expected soon~\cite{Serguey}.

\subsection{SME: Neutrino sector}
\label{sec:neutrino-const}
Neutrino physics can be affected in several ways by Lorentz breaking operators. The most common channels for casting constraints are observations concerning the speed of neutrinos with respect to light, modified oscillations between neutrinos flavors and threshold reactions.

\paragraph{Constraints from neutrino time of flights:} In assessing the limit speed of neutrinos with respect to that of light, we
have to date only a single event to rely on, the supernova SN1987a. This was a peculiar event which allowed to detect the almost simultaneous (within a few hours) arrival of electronic antineutrinos and photons. Although only few electronic antineutrinos at MeV energies was detected by the experiments KamiokaII, IMB and Baksan, it was enough to establish a constraint $(\Delta c/c)^{TOF} \lesssim 10^{-8}$ \cite{Stodolsky:1987vd} or $(\Delta c/c)^{TOF} \lesssim 2\times10^{-9}$ \cite{Longo:1987gc} by looking at the difference in arrival time between antineutrinos and optical photons over a baseline distance of $1.5\times10^5$ ly. Further analyses of the time structure of the neutrino signal strengthened this constraint down to $\sim10^{-10}$ \cite{Ellis:2008fc,Sakharov:2009sh}. The scarcity of the detected neutrino did not allow the reconstruction of the full energy spectrum and of its time evolution in this sense one should probably consider constraints purely based on the difference in the arrival time with respect to photons more conservative and robust. Unfortunately adopting $\Delta c/c \lesssim 10^{-8}$, the SN constraint implies very weak constraints, $\xi_{\nu}^{(3)}\lesssim 10^{13}$ and $\xi_{\nu}^{(4)}\lesssim 10^{34}$. Note that observational constraints on $\Delta c/c$ translate in constraints on the LIV parameter via the formula~\cite{Maccione:2011fr} 
\begin{equation} 
\xi^{(n)}_{\nu}= \frac{2}{n-1}\left(\left.\frac{\Delta c}{c}\right|^{TOF}_{\rm Obs}+ \frac{m^2}{2p^2}\right)\times\left(\frac{\Mpl}{p}\right)^{(n-2)} 
\;.
\label{eq:mtof}
\end{equation}

\paragraph{Constraints from neutrino oscillations:} At odd with the previous case, we do have a wealth of information only about neutrino oscillations which however constraints only the differences among LIV coefficients of different flavors. The best constraint to date comes from survival of atmospheric muon neutrinos observed by the former IceCube detector AMANDA-II in the energy range 100 GeV to 10 TeV \cite{Kelley:2009zza}, which searched for a generic LIV in the neutrino sector~\cite{GonzalezGarcia:2004wg} and achieved $(\Delta c /c)_{ij} \leq 2.8\times10^{-27}$ at 90\% confidence level assuming maximal mixing for some of the combinations $i,j$. The same constraint applies to the corresponding antiparticles as IceCube does not distinguish neutrinos from antineutrinos. IceCube is also expected to improve this constraint to $(\Delta c / c)_{ij} \leq 9\times 10^{-28}$ in the next few years \cite{Huelsnitz:2009zz}. The lack of sidereal variations in the atmospheric neutrino flux also yields comparable constraints on some combinations of SME parameters \cite{Abbasi:2010kx}. Putting all together, it seems that no-flavour dependent SME can be tolerated and many studies just assume flavour independent LIV as they starting base of their analysis.

\paragraph{Constraints from neutrino threshold reactions:} Several threshold processes have been considered in the literature, most prominently the neutrino \v{C}erekov emission $\nu\to\gamma\,\nu$, the neutrino splitting $\nu\to \nu\,\nu \overline{\nu}$ and the neutrino electron/positron pair production $\nu\to\nu \,e^{-}e^{+}$. They are all very similar, so we shall discuss here only the latter for illustrational purposes and assume no LIV modification in the electron/positron sector as we have already seen that LIV in this sector is strongly constrained. 
The threshold energy is for arbitrary $n$ is then
\begin{equation}
E_{th, (n)}^{2} = \frac{4m_{e}^{2}}{\delta_{(n)}} \qquad  \mbox{with} \quad\delta_{(n)} = \xi^{(n)}_{\nu}\left(\frac{E_{th,(n)}}{M_{\rm Pl}}\right)^{n-2}\; .
\end{equation}

The rate of this reaction was firstly computed in \cite{Cohen:2011hx} for $n=2$ but can be easily generated to arbitrary $n$~\cite{Maccione:2011fr} (see also \cite{Carmona:2012tp} for a more general treatment and detailed considerations). The generic energy loss time-scale then reads (dropping purely numerical factors)

\begin{equation}
\tau_{\nu{\rm -pair}} \simeq \frac{ m_Z^4 \cos^4 \theta_w}{g^4E^5} \left( \frac {M_{\rm Pl}} {E} \right)^{3(n-2)}\;,
\label{eq:tau-nu-pair}
\end{equation}
where $g$ is the weak coupling and $\theta_w$ is Weinberg's angle.

The observation of upward-going atmospheric neutrinos up to 400 TeV by the experiment IceCube implies that the free path of these particles is at least longer than the Earth radius implies a constraint $\xi^{(3)}_\mu\lesssim 40$ (taking a conservative baseline of about 6000 Km). No effective constraint can be obtained for $n=4$ LIV, however in this case neutrino splitting (which has the further advantage to be purely dependent on neutrinos LIV) could be used on the ``cosmogenic'' neutrino flux which one expects from the decay of charged pions produced by the aforementioned GZK reaction. 

It is easy to see that the neutrino splitting should modify the spectrum of the ultra high energy neutrinos by suppressing the flux at the highest energies and enhancing it at the lowest ones. In \cite{Mattingly:2009jf} it was shown that future experiments like ARIANNA \cite{Barwick:2006tg} will achieve the required sensitivity to cast a constraint of order $\xi^{(4)}_{\nu} \lesssim 10^{-4}$. Note however, that the rate for neutrino splitting computed in \cite{Mattingly:2009jf} was recently recognised to be  underestimated by a factor $O(E/M)^2$~\cite{Ward:2012fy}. An improved analysis can be found in ~\cite{Maccione:2011fr} leading to an expected constraint from AUGER $\xi^{(4)}_{\nu} \lesssim 10^{-7}$ (note that this constraint is cast using the expected flux which won't be able to distinguish different flavours).

However, we have just seen that experimental observations of the depth of the shower maximum of UHECR interactions in the atmosphere hinted at the possible presence of nuclei heavier than protons in UHECRs \cite{Abraham:2010yv}. In this case, pion production would be suppressed at UHE and hence  the UHE neutrino flux could be much smaller than the expectation from pure proton composition.

A final comment is devoted to the so called pion decay channel $\pi^+\to \nu_\mu \mu^+$ that was extensively explored in dealing with the recent OPERA claim~\cite{Bi:2011nd,Cowsik:2012qm}. In this case it was found that for superluminal neutrinos and unmodified pion and muon, the process could be actually forbidden hence suppressing muonic neutrino pair production in the CERN bean. In ~\cite{Bi:2011nd} it was noticed that the detection of up to 40 GeV neutrinos at OPERA would imply a bound $\xi^{(2)}_{\nu} \lesssim 10^{-7}$. Also in~\cite{Bi:2011nd,Cowsik:2012qm} a constraint $\xi^{(2)}_{\nu} \lesssim 10^{-13}$ was derived from atmospheric neutrinos using neutrino pair production and neutrino \v{C}erenkov emission. 

While OPERA detection of superluminal neutrinos was soon recognised as flawed, it is worth reconsider it here as a case study. This will also illustrate some of the pitfalls that may be encountered when applying the above mentioned techniques to specific experimental situations.

\section{The Opera affaire: what was said and what could have been said}

In fall 2011 the OPERA collaboration's reported \cite{Adam:2011zb} and then retracted (see revised version of \cite{Adam:2011zb}) a measurement of the time of arrival of non-oscillated $\nu_\mu$ over the path length $\sim730$km from CERN to Gran Sasso with average energy of $\sim17$~GeV which was in open conflict with Special Relativity.
The $\nu_\mu$ seemed to arrive earlier than light would, by an amount $
\Delta t = 60.7 \pm 6.9\, {\rm (stat.)} \pm 7.4\, {\rm (sys.)}$ corresponding to an apparent propagation velocity of ${\Delta c}/{c} = (2.48 \pm 0.28\, {\rm (stat.)} \pm 0.30\, {\rm (sys.)})\times10^{-5}$ where $c$ is the low energy speed of light in vacuum, and $\Delta c = v_\nu - c$. 
After discovering a flaw in the initial measurement and the realisation of a new one, the OPERA result (${\Delta c}/{c} = (2.7 \pm 3.1\, {\rm (stat.)} \pm 3.4\, {\rm (sys.)})\times10^{-6}$), together with the additional results of Icarus \cite{Antonello:2012be}, and the MINOS Collaboration analysis \cite{Adamson:2007zzb} now can be seen as constraints on neutrino velocities in the GeV range rather than a signal of beyond standard model physics.

The excitement created by the OPERA's initial report has obviously subsided for the greater physics community. For the quantum gravity community that focuses on possible experimental signatures of quantum gravity, however, technical issues were raised in how to analyse these types of accelerator based experiments properly.  In particular, the detailed physics of anomalous reactions, and how they reduce the intensity of a particle beam from source to detector became central to the discussion.  

The most convincing theoretical objection to the initial OPERA result was produced by Cohen and Glashow~\cite{Cohen:2011hx} shortly after the OPERA report and involved just such an anomalous reaction.  Cohen and Glashow used the fact that superluminal neutrinos should emit electron-positron pairs (see section~\ref{sec:neutrino-const}) to argue that the OPERA results were not even self-consistent: any neutrino with the speed reported by OPERA should have lost most of its energy to pair production while it propagated from CERN to the detector at Gran Sasso. The maximum energy in the beam would therefore have dropped to be below some termination energy $E_T$, and Cohen and Glashow showed that $E_T$ for the OPERA beam was less than the average $\sim17$~GeV energy reported by OPERA.  

More specifically by integrating the energy loss rate from pair production (as deducible from Eq.~(\ref{eq:tau-nu-pair})) over a distance  $L$ and by assuming that the typical energy loss length be much smaller than $L$ one obtains
\begin{equation} \label{eq:termE}
E^{-3n+1} - E_{0}^{-3n+1} = (3n-1)\xi_{\nu}^{3}E_{\rm ref}^{-3(n-2)}k \frac{G_{F}^{2}}{192\pi^{3}}L \equiv E_{T}^{-3n+1}\;,
\end{equation}
where $E$ is the energy on a neutrino starting with energy $E_{0}$ after propagation over the distance $L$  and $E_{\rm ref}$ is the energy at which we normalize the parameter $\xi_{\nu}$. The factor $k=25/448$ was computed in \cite{Cohen:2011hx} for the case $n=2$, while for the general case it can be found in \cite{Carmona:2012tp}.  The ``termination'' energy $E_{T}$ corresponds to the energy that a neutrino would approach after propagation over a distance $L$ if it started with $E_{0}\gg E_{T}$. We remark here that the termination energy $E_{T}$ is a mildly varying function of $n$ and of the energy scale $E_{\rm ref}$. For a LIV extension of the standard model at order $n=2$ (see the Coleman-Glashow model discussed briefly above) it was then shown that the Opera claim of $\xi_{\nu} \sim 5\times10^{-5}$ for $E_{\rm ref}\sim 10~\mathrm{to} ~30$~GeV implied $E_{T}\simeq 12.5$~GeV: a value of $E_{T}$ obviously incompatible with the observation of neutrinos up and above 40 GeV in Opera. Similar considerations lead to an expectation of $E_{T}\simeq 15.0$~GeV for $n=3$.

Note that this estimate was done without taking into account the similar mechanism of neutrino splitting. While this reaction is strictly forbidden in the $n=2$ case, it is not so at higher orders and needs to be taken into account given that the energy loss rate of this process is comparable to the one for pair production loss  (see e.g.~\cite{Maccione:2011fr,Ward:2012fy,Carmona:2012tp}), and hence is not negligible for $n>2$. In particular, this extra reaction can lead to energy losses without generating large numbers or electron-positron pairs as those searched for in an early ICARUS analysis of the Opera beam~\cite{Antonello:2012hg}.

The physics of the Cohen-Glashow argument was correct, however the authors did not worry about adjusting for the finite size of the baseline. A finite baseline can be of the same order as the energy loss length of neutrinos undergoing pair production. This allows for some neutrinos to undergo only one or a few Cherenkov emissions within their time of flight. Therefore the most energetic neutrinos of the injection beam can still  reach the end of the baseline with an energy larger than $E_{T}$~\cite{Maccione:2011fr}. It is then necessary, in order to cast a robust constraint on LIV by using long baseline experiments, to run a full MonteCarlo simulation of the propagation of neutrinos aimed at computing the neutrino spectrum on arrival in the presence of this energy loss process.  While this was not an issue for the Cohen and Glashow result, as it was one piece of a number of experimental and theoretical concerns about OPERA~\cite{Antonello:2012be,Antonello:2012hg,Stodolsky:1987vd,Longo:1987gc}, if one wishes to use time of flight experiments alone to set robust constraints on neutrino LIV, the issue must be addressed.

In~\cite{Maccione:2011fr} a complete analysis for the case of OPERA in the cases $n=2$ and $n=3$, taking into account neutrino pair creation and neutrino splitting (for $n=3$) has been performed. It was there found that the propagated spectrum does indeed show a pronounced bump at the expected $E_{T}$, but is also characterised by a high energy tail that extends well above $E_{T}$ and has an amplitude about 10\% of the amplitude of the bump. 
Hence, the simple calculation of $E_{T}$ is not per se conclusive for casting constraints, although the reconstruction of the propagated spectra in~\cite{Maccione:2011fr} demonstrated that in the special case of OPERA the detection of neutrinos with $E>40$~GeV would have still pointed out an incompatibility between the adopted LIV framework and the experimental observation.

\section{From LIV EFT to Quantum gravity phenomenology}

So far we have reviewed several constraints casted on a very conservative dynamical framework incorporating systematically departures from Lorentz invariance. Of course this has an informative value per se, given that many QG models are expected to admit an EFT limit well below the Planck scale.

Nonetheless, it is also clear that the outreach of our constraints is somewhat limited by the very same generality of our framework and that whenever possible it should be now time to attempt the bolder step to apply the very same techniques developed in EFT to specific QG frameworks or observational/experimental evidence.

In this sense we shall now consider a specific application of the above mentioned ideas with the aim to show how the past expertise could be sometime applied effectively to QG candidate models and observations. We shall start by first introducing an analogue of the mSME for the gravitational sector, i.e. the so called Einstein--Aether theory of gravity with its current constraints. Then we shall introduce a candidate QG theory which might admit a restricted version of Einstein--Aether as an IR limit and discuss how this can be constrained using the previous techniques, stressing in this way the far-fetching implications that the so-obtained constraints can have on the theory.

\subsection{Einstein--Aether theory}

The SME is constructed by coupling matter terms to non-zero LIV tensors in vacuum.  However, when dealing with extensions of this framework to gravitational phenomena it is clear that leaving this tensors non-dynamical, would break general covariance.  In this respect one can proceed along two alternative approaches, the first being the introduction of a suitable dynamics for the LIV tensor, the other being the the acceptance of an explicit breaking of the four dimensional diffeomorphism invariance of GR. In what follows we shall discuss the most representative models in this respect and they relation.

If one choses to preserve general covariance by promoting the LIV tensors to dynamical fields and if restricts his/her attention to rotational invariance, then it is natural to generate LIV couplings by including in the action either a scalar or a timelike vector field that takes a vacuum expectation value.  In the case of a scalar, one can use a shift symmetry ($\phi(x) \rightarrow \phi(x) + \phi_0$) to construct actions for which the derivative of the scalar takes a non-zero value~\cite{ArkaniHamed:2003uy}.  In the vector case, one simply puts a potential for the vector field such that the vector acquires a vev.  

We concentrate on the vector case here as it is the simplest model that allows for rotationally invariant Lorentz violation~\cite{Jacobson:2000xp} and it is also the natural extension of the framework considered so far for the rotationally invariant LIV matter sector.  It is the most general theory for a unit timelike vector field coupled to gravity (but not to matter), which is second order in derivatives. 

Let us the express again the aether vector field by $u^\alpha$ and, in analogy with what we did for the SME, write  the most general theory for a unit timelike vector field coupled to gravity. If we limit ourself to second order terms in derivatives (or equivalently low energies) this take the form
\begin{equation}\label{ac:ae}
{\cal S} = {\cal S}_{EH} + {\cal S}_{u} = \frac{1}{16\pi\Gae}\int d^{4}x\sqrt{-{g}}\; \left(R + {\cal L}_{u}\right).
\end{equation}
where ${\cal L}_{u}$ is given by
	\begin{equation}\label{lag:ae}
	{\cal L}_{u} = -\tensor{Z}{^{\alpha \beta}_{\gamma \delta}}(\Dl_\alpha u^\gamma)(\Dl_\beta u^\delta) + V(u).
	\end{equation}
The tensor ${Z}{^{\alpha \beta}_{\gamma \delta}}$ is defined as~\cite{Jacobson:2008aj}
	\begin{equation}\label{def:Zabcd}
	\tensor{Z}{^{\alpha \beta}_{\gamma \delta}} = c_1 g^{\alpha \beta}g_{\gamma \delta} + c_2\tn{\delta}{^\alpha_\gamma}\tn{\delta}{^\beta_\delta} + c_3\tensor{\delta}{^\alpha_\delta}\tensor{\delta}{^\beta_\gamma} - c_4u^\alpha u^\beta g_{\gamma \delta}~,
	\end{equation}
where $c_i,\,i = 1,\ldots,4$ are simple coefficients of the various kinetic terms and $V(u)$ is a potential term such that it generates a non-zero vev for $u^\alpha$.  Note the indicial symmetry $\tn{Z}{^{\beta \alpha}_{\delta \gamma}} = \tensor{Z}{^{\alpha \beta}_{\gamma \delta}}$. An additional term, $R_{ab} u^a u^b$ is a combination of the above terms when integrated by parts, and hence is not explicitly included here.

The potential in (\ref{lag:ae}) is normally fixed by the requirement to remove the ghost excitation associated to the one of the vector components which will necessarily acquire a wrong sign in the kinetic term. The choice $V(u)=\lambda(u^2+1)$, where $\lambda$ is a Lagrange multiplier,  fixes the norm of $u^\alpha$ and removes the ghost excitation (c.f. the discussions in \cite{Eling:2004dk} and \cite{Bluhm:2008yt}).  It is this choice of the potential that it is normally associated to the so called ``Einstein--Aether theory''~\cite{Jacobson:2000xp} which we shall consider here.

\subsubsection{Constraints on Einstein--Aether gravity:}

For what regards the constraints on Einstein--Aether gravity, they can be divided in those on the aether kinetic terms and those on the {ae}ther-matter couplings. Given the the latters can be reduced to mSME constraints, we shall here discuss specifically only the first kind (one can find a discussion of the second kind in e.g.~\cite{Liberati:2012th}). 

Before of doing so, we can note however that the couplings between aether and the standard model field content while being the same as those discussed before in this review, they do have new features as the aether field has now dynamics, so there can be position dependent violations of Lorentz symmetry.  Interestingly, some of the couplings to matter that are unobservable for a single fermion field can have relevant effects when the aether varies.  For example, the $-a u_\mu \overline{\psi} \gamma^\mu \psi$  term in the mSME could be removed by making a phase change for the fermion.  However, once $u^\alpha$ is dynamical and varies with position, only a single component of the term can actually be removed by a phase change~\cite{Kostelecky:2008in}.  This leads to a new type of term which requires gravitational/position dependent tests in the matter sector~\cite{Kostelecky:2008in}.  

%
%

In order to constrain instead {ae}ther kinetic terms one can adopt two different approaches. One consists of course in adopting, as normally done any modified gravity theory, a PPN analysis that allows to compare the theory with observations e.g. solar system constraints. Alternatively, one can use the fact that the theory predicts extra degrees of freedom in the gravity sector (there are naively four, but the unit constraint removes one) and that these excitations strongly couple to the metric via the unit constraint (for a more detailed discussion see \cite{Jacobson:2008aj, Liberati:2012th}). 

\paragraph{Constraints from PPN analysis:} All the PPN parameters vanish except for $\alpha_1,\alpha_2$ which describe preferred frame effects.  $\alpha_1$ and $\alpha_2$ were calculated in~\cite{Foster:2005dk}
\begin{eqnarray}
\alpha_1=\frac {-8(c_3^2+c_1c_4)} {2c_1-c_1^2+c_3^2}\\
\alpha_2=\frac {\alpha_1} {2}- \frac {(c_1+2c_3-c_4)(2c_1+3c_2+c_3+c_4)} {(c_1+c_2+c_3)(2-c_1-c_4)}.
\end{eqnarray}
Current constraints are $\alpha_1<10^{-4}$ and $\alpha_2<4 \times 10^{-7}$~\cite{Will:2005va} and so from a PPN analysis alone there is still a large 2-d region of parameter space that remains consistent with available tests of GR.

\paragraph{Constraints from gravity-aether wave modes:} The combined aether-metric modes consist of the two usual transverse traceless graviton modes, a vector mode, and a scalar mode~\cite{Jacobson:2004ts}.   The speeds of each of the modes can differ from the speed of light.  Hence if the speeds are less than unity, high energy cosmic rays will emit vacuum gravitational \v{C}erenkov radiation~\cite{Elliott:2005va}.  If we denote the speeds of the spin-2, spin-1 and spin-0 modes by $s_2, s_1, s_0$ then we have~\cite{Jacobson:2004ts}
\begin{eqnarray}
s_2^2=(1-c_1-c_3)^{-1}\\
s_1^2=\frac{2c_1-c_1^2+c_3^2} {2(c_1+c_4)(1-c_1-c_3)}\\
s_0^2=\frac{(c_1+c_2+c_3)(2-c_1-c_4)}{(c_1+c_4)(1-c_1-c_3)(2+3c_2+c_1+c_3)}.
\end{eqnarray}
The requirement that all these speeds are greater than unity therefore puts constraints on a combination of the $c_i$ coefficients.
However, even after imposing all of the above constraints there is still a large region of parameter space allowed.  Indeed, the PPN and gravitational \v{C}erenkov constraints are all satisfied provided quite lose conditions on the model coefficients are satisfied~\cite{Jacobson:2008aj,Liberati:2012th}. Hence, the gravitational sector is only minimally constrained compared to aether-matter couplings.

\subsection{Ho\v rava--Lifshitz  gravity}
\label{sec:horava-int}

The underlying idea of the Ho\v rava--Lifshitz (HL) gravity (see {\em e.g.~}\cite{Sotiriou:2010wn} for a review) is to achieve power-counting renormalizability by modifying the graviton propagator in the ultraviolet by adding to the action terms containing higher order spatial derivatives of the metric, but not higher order time derivatives, so to preserve unitarity.   This procedure naturally leads to a space-time foliation into spacelike surfaces, labeled by the $t$ coordinate and with $x^i$ being the coordinates on each surface. The resulting theory is then invariant only under the reduced set of diffeomorphisms that leave this foliation intact, $t\to \tilde{t}(t)$ and $x^{i}\to \tilde{x}^{i}(t,x^i)$. 

It was shown that power counting renormalizability requires the action to includes terms with at least 6 spatial derivatives in 4 dimensions \cite{Horava:2009uw,Visser:2009fg}. Of course, all lower order operators compatible with the symmetry of the theory are expected to be generated by radiative corrections, so the most general action takes the form \cite{Blas:2009qj}
\begin{equation}
\label{SBPSHfull}
S_{HL}= \frac{M_{\rm Pl}^{2}}{2}\int dt d^3x \, N\sqrt{h}\left(L_2+\frac{1}{M_\star^2}L_4+\frac{1}{M_\star^4}L_6\right)\,,
\end{equation}
where $h$ is the determinant of the induced metric $h_{ij}$ on the spacelike hypersurfaces,
\begin{equation}
L_2=K_{ij}K^{ij} - \lambda K^2 
+ \xi {}^{(3)}\!R + \eta a_ia^i\,,
\end{equation}
where $K$ is the trace of the extrinsic curvature $K_{ij}$, ${}^{(3)}\!R$ is the Ricci scalar of $h_{ij}$, $N$ is the lapse function, and $a_i=\partial_i \ln N$. 
$L_4$ and $L_6$ denote a collection of 4th and 6th order operators respectively and $M_\star$ is the scale that suppresses these operators which does not coincide {\em a priori} with $M_{\rm Pl}$. 

It is perhaps tempting to call $M_\star$ the Lorentz breaking scale, but the theory exhibits Lorentz violations (LIV) at all scales, as $L_2$ already contains LIV operators. These Infrared (IR) Lorentz violations are controlled by three dimensionless parameters that take the values $\lambda=1$, $\xi=1$ and $\eta=0$ in General Relativity (GR). While, $\xi$ can be set to 1 by a suitable coordinate rescaling, it is presently unclear if the running of the remaining two parameters will converge on the GR values in the IR. Nonetheless, it seems that the the theory could still be viable and consistent for suitable choices of the dimensionless parameters $\lambda$, $\eta$ which admits the GR values as extremal limits of the allowed range \cite{Blas:2010hb}. 

Action (\ref{SBPSHfull}) does present, however, the unappealing feature to contain (in $L_4$ and $L_6$)  a very large number, $O(10^2)$, of operators and independent coupling parameters. In remedy of this situation, restrictions to the theory have been proposed which would limit the proliferation of independent couplings. We shall not deal with such restrictions here (but see e.g~\cite{Sotiriou:2010wn,Vernieri:2011aa} for a concise review) as they will not be determinant for the phenomenological discussion on HL that we shall present later on in this review.

\subsection{Relation between Einstein--Aether and Ho\v rava--Lifshitz gravity}

One interesting, and at the same time problematic, feature of HL gravity is the presence of a new propagating scalar mode associated with the reduced diffeomorphism invariance of the theory with respect to GR.  However, if one chooses to restore diffeomorphism invariance, then this mode manifests as a foliation-defining scalar field \cite{Jacobson:2010mx}. This field allows also to make manifest the relation between the $L_2$  Lagrangian and the Einstein--Aether one.

In fact, it was shown in \cite{Jacobson:2010mx} (but see also~\cite{Afshordi:2009tt}) that the Einstein--Aether theory is equivalent to the the infrared limit of HL gravity if the aether is assumed to be hypersurface orthogonal before the variation. More precisely, hypersurface orthogonality can be imposed through the local condition
\begin{equation}
u_\mu=\frac{\partial_\mu T} {\sqrt{g^\alpha\beta \partial_\alpha T \partial_\beta T}}\, ,
\end{equation}
where $T$ is a scalar field that defines a foliation. Choosing $T$ as the time coordinate one selects the preferred foliation of HL gravity, and the action (\ref{ac:ae}) reduces to the action of the infrared limit of HL gravity, whose Lagrangian we denoted as $L_2$ in eq. (\ref{SBPSHfull}). The details of the equivalence of the equations of motions and the correspondence of the parameters of the two theories is discussed in detail in \cite{Barausse:2012ny,Barausse:2012qh}.

As a concluding remark let us stress that the Einstein--Aether theory can be seen as a sort of gravity sector version of the rotational invariant mSME. However, one should expect to be able to supplement also Einstein--Aether gravity with suitable higher order operators. This has not been done yet, but the existence of HL gravity and its relation in the IR with Einstein--Aether theory seems to suggest that such an extension should be viable. For example, a possible extension of HL gravity to a theory invariant under the full group of four-dimensional diffeomorphisms have been proposed~\cite{Germani:2009yt}. This kind of extensions could then be seen as possible generalisations of the Einstein--Aether theory beyond the IR limit.

\subsection{Constraints on Ho\v{r}ava gravity}
\label{sec:horava}

Coming to constraints on HL gravity theory, it should be obvious that given the relation in the IR between hypersuface orthogonal Einstein--{Ae}ther and Ho\v{r}ava--Lifshiftz gravity, the previously presented constraints on Einstein--Aether gravity can in principle be related to constraints for the latter theory. 

Looking then at the UV complete the theory from the point of view of QG phenomenology, it then interesting to know the available constraints on the Lorentz breaking scale of the theory $M_\star$. Remarkably, this scale happens to be bounded both below and above. Indeed, for the theory to preserve power counting renormalizablity and be at the same time compatible with current observations (microgravity experiments and solar system tests)  one has to require $O(1)\,{\rm meV} <M_\star < 10^{16}\,\, {\rm GeV}$~\cite{Blas:2010hb,Sotiriou:2010wn,Liberati:2012jf}, a quite broad opportunity window for the theory.

We have seen however, that radiative corrections will always allow for LIV operators to percolate from a SM sector to the other, gravity being no exception. In this sense it can be generically expected for the above theory to induce Lorentz breaking operators in the matter sector at all orders. Let us assume here that no LIV is present in the matter sector at tree level and that again some protective mechanism will prevent the percolation of the Lorentz breaking terms to the lowest order (mass dimension 3 and 4) operators of the matter sector (we shall come back later on this point). Also, one can assume that the CPT and Parity (P) invariance of the gravitational action is preserved in the matter sector. Indeed if no CPT and P odd operators are present in the matter sector at the tree level one would not expect them to be generated via radiative corrections induced by the gravitational, CPT and P even, terms. This assumption forbids helicity dependent terms and allows only even power of the momentum in the matter dispersion relation. Hence matter and anti-matter are expected to share the same dispersion relation which within this framework one can then expect to be
 \begin{equation}
E^{2} = m^{2} + p^{2} + \eta \frac{p^{4}} {M_{\rm LIV}^{2}} +O\left(\frac{p^{6}}{M_{\rm LIV}^{4}} \right)\, .
\label{eq:mattHL-disp-rel}
\end{equation}

From a logic point of view, there are  now two options: (a) $M_\star=M_{\rm LIV}$, i.e.~$M_\star$ is a universal scale; (b) $M_\star\ll M_{\rm LIV}$ (as we have seen in this review current phenomenological constraints already rule out the case $M_{\rm LIV} \ll M_{\star}$). Clearly, option (b) requires some mechanism which suppresses the percolation of LIVs in the matter sector even to higher order operators. We already discussed such a mechanism, the so called ``gravitational confinement" \cite{Pospelov:2010mp}.  Here, reporting the work done in~\cite{Liberati:2012jf}, we shall conclude the necessity for such mechanisms. 

For doing so the strategy adopted in \cite{Liberati:2012jf} consisted in focussing on option (a) and demonstrating that, in this case, matter LIV constraints imply $M_\star=M_{\rm LIV}>10^{16}$~GeV, thus closing the available window for $M_\star$. Of course this could be easily achieved using the constraints on QED for modified dispersion relation of order  $n=4$ using UHECR. However, we have seen that these constrains are somewhat questionable nowadays while further evidence about the nature of the highest energy particles is awaited. 

Hence, in \cite{Liberati:2012jf}, it was suggested to use the more robust observation of the synchrotron radiation from the Crab Nebula. A first estimate of the strength of the available constraint can be derived along the lines described before. In particular for a negative $\eta$ one can run the argument that the maximal obtainable frequency (\ref{eq:ommax}) should not be smaller than the observed maximal one.

The maximal observed frequency in the CN synchrotron spectrum, $\omega_{\rm obs}\approx 0.1$~GeV,  for $n=4$ one can then analytically derive a constraint $\eta \gtrsim -10^{5}$ (assuming $B\sim300~\mu$G and $M_{\rm LIV} = M_{\rm pl}$), which would correspond to $M_{\rm LIV} > 3\times10^{16}$~GeV. 

While promising, this is not a double sided constraint. In order to be sensitive to the full range of the $\eta$ parameters and take into account competing LIV and LI effects a much deeper analysis is needed. As we saw, this was performed in \cite{Maccione:2007yc} where the possible LIV induced modifications to the standard Fermi mechanism (which is thought to be responsible for the formation of the spectrum of energetic electrons in the CN) were considered and the synchrotron spectrum of the CN was recomputed taking into account all the new, LIV induced, phenomena for $n=3$ LIV. This was done in~\cite{Liberati:2012jf} for $n=4$ and the specific modified dispersion relation Eq.~(\ref{eq:mattHL-disp-rel}). 

The free parameters of the model (electron/positron density and spectrum and magnetic field strength) were fixed in order to reproduce the low energy part of the spectrum, which is not affected by LIV 
This allows to reconstruct how LIV affects the higher energy part of the spectrum ($E\gtrsim100$~keV) and to use a $\chi^{2}$ statistics to measure when deviations from the observed spectrum due to LIV become unacceptably large. 
%
By considering the offset from the minimum of the reduced $\chi^{2}$ exclusion limits at 90\%, 95\% and 99\% Confidence Level (CL), according to \cite{pdg} it was shown in \cite{Liberati:2012jf} that mass scales $M_{\rm LIV} \lesssim 2\times10^{16}$~GeV are excluded at 95\% CL.

So one can conclude that the application of the previously described methods to concrete QG model leads to a couple of important lesson. First of all, one should not think a priori that some methods are effective only for some order of LIV EFT. The synchrotron constraint we considered here is apparently ineffective for $n=4$ and $M_{\rm LIV}=M_{\rm Pl}$, however it is not obvious that any QG model should send the LIV scale equal to the Planck one. 
Secondly, the constraint obtained has broader implication for the model that could not be foreseen in a  purely EFT approach. In particular it implies necessarily $M_* \ll M_{\rm LIV}$ and hence that a protective mechanism beyond the lowest order operator should be foreseen in the matter sector of Ho\v{r}ava gravity. As we discussed previously  it has been shown that dimension 4 operators of matter can be efficiently screened from LIV this way in HL models \cite{Pospelov:2010mp} and this is expected to be the case for higher order operators as well. The investigation we just reported can then be taken as a strong indication that such a mechanism should be considered the main avenue for the making this particular QG proposal viable.

\section{Beyond EFT: other frameworks}
\label{sec:others}

Specifying which dynamical framework is employed is crucial when discussing the phenomenology of Lorentz violations. In this review we have focussed on the the most conservative framework, effective field theory.  However, this is not the only one conceivable and, in fact, there are reasonable arguments from holography that a quantum gravity theory should not necessarily be a local field theory in the UV (c.f. the discussion in~\cite{Shomer:2007vq}).  Hence deviation from standard Lorentz invariance may enter into low energy physics in novel ways.  Given that the EFT approach is nothing more than a highly reasonable, but rather arbitrary ``assumption'', it is worth studying and constraining additional models, given that they may evade the majority of the constraints discussed in this review.



\subsection{D-brane models}
\label{sec:nonEFT}

A class of models showing modified dispersion relations of the general form Eq.~(\ref{eq:disprel}) was derived from the Liouville string approach to quantum space-time \cite{Ellis:1992eh,Ellis:1999jf,Ellis:2003sd}. These models motivate corrections to the usual relativistic dispersion relations that are first order in the particle energies and that correspond to a vacuum refractive index $\eta = 1-(E/M_{\rm Pl} )^{\alpha}$, where $\alpha = 1$. Models with quadratic dependences of the vacuum refractive index on energy: $\alpha = 2$ have also been considered \cite{Burgess:2002tb}. 

Most importantly, the D-particle realisation of the Liouville string approach predicts that only gauge bosons such as photons, and not charged matter particles such as electrons, might have QG-modified dispersion relations. This occurs since excitations which are charged under the gauge group are represented by open strings with their ends attached to the D-brane \cite{Polchinski:1996na}, and only neutral excitations are allowed to propagate in the bulk space transverse to the brane \cite{Ellis:2003if}. Thus, if we consider photons and electrons, in this model the parameter $\eta$ is forced to be null, whereas $\xi$ is free to vary. Noticeably, the theory is also by construction CPT even, implying that vacuum is not birefringent for photons ($\xi_{+} = \xi_{-}$). As such these models are very difficult to constraint. However, one of the possible incarnations of this model with $\alpha=1$ was constrained in~\cite{Maccione:2010sv} by using again the absence of upper threshold for secondary photons of the GZK reaction. The obtained constraint on the photon LIV parameter is of order $O(10^{-12})$ however alternative spacetime foam models for which this constraint does not apply can be envisaged~~\cite{Ellis:2010he}.

\subsection{New Relativity Theories}

An alternative route with respect to explicit Lorentz breaking pursued in the extant literature consists in trying to define a new relativity group which would allow for an extra invariant scale (generally the Planck length) beyond the speed of light. In this direction is then worth exploring generalisation of the von Ignatowski derivation which would keep the relativity principle while relaxing one or more of the above discussed postulates. 

For example, relaxing the space isotropy postulate leads to the so-called Very Special Relativity framework \cite{Cohen:2006ky}, which was later on understood to be associated to a Finsler-type geometry~\cite{Bogoslovsky:2005cs,Bogoslovsky:2005gs,gibbons:081701}. In this example, however, the generators of the new relativity group number fewer than the usual ten associated with Poincar\'e invariance. Specifically, there is an explicit breaking of the $O(3)$ group associated with rotational invariance. 

Much research has instead focus on constructing alternative relativity groups with the same number of generators as special relativity. Currently, we know of no such generalisation in coordinate space within the standard  commutative geometry framework. However, it has been suggested that, at least in momentum space, such a generalization is possible, and it was termed  ``doubly" or ``deformed" (to stress the fact that it still has 10 generators) special relativity (DSR)~\cite{AmelinoCamelia:2000mn, Magueijo:2001cr,Magueijo:2002am, AmelinoCamelia:2002gv}.  Such momentum space construction have been linked to possible Lie-type non commutative geometry (mainly the so called kappa-Minkowski one~\cite{Freidel:2003sp,Freidel:2007yu}) or to curved momentum spaces in higher dimensions~\cite{Girelli:2006mg}. Even though DSR aims at consistently including dynamics, a complete formulation capable of doing so is still missing, and present attempts face major problems. 

Also to overcome some of these conceptual problems was recently advanced what we might call a ``spin-off" of DSR named Relative Locality. This framework is based on the idea that the invariant arena for classical physics is a curved momentum space rather than spacetime (the latter being a derived concept)~\cite{AmelinoCamelia:2011bm,AmelinoCamelia:2011pe,Carmona:2011wc,AmelinoCamelia:2011yi,KowalskiGlikman:2012ji}. Interestingly, while this seems to allow an extension of the relativity principle with an invariant energy scale, it does so at the price to renounce to an observer-independent concept of locality. There is some evidence that this might be generic feature of any alternative relativity group of this sort \cite{Carmona:2011wc}, suggesting locality violations in EFT as a possible new avenue of exploration in QG phenomenology (see also discussion on non-locality in QG below).

In summary, DSR and Relative Locality are still a subject of active research and debate (see e.g.~\cite{Smolin:2008hd,Rovelli:2008cj,AmelinoCamelia:2011uk,Hossenfelder:2012vk}); nonetheless, they are reaching just now the level of maturity required for casting constraints (see e.g.~\cite{AmelinoCamelia:2011bq}). 

\section{Discussion and perspectives}

We summarize the current status of the constraints for the LIV SME (rotational invariant) in Table~\ref{table:sum}.
\begin{table}[!htb]
\begin{center}
\footnotesize
\centering
\begin{tabular}{p{1.2cm}|p{2.7cm}|p{2.7cm}|p{2.7cm}|p{2.7cm}}
\hline\noalign{\smallskip}
Order & photon & $e^{-}/e^{+}$ & Protrons & Neutrinos$^a$  \\
\noalign{\smallskip}\hline\noalign{\smallskip}
n=2 & N.A. & $O(10^{-16})$ & $O(10^{-20})$ (CR) & $O(10^{-8}\div10^{-10})$ \\
n=3 & $O(10^{-16})$ (GRB) & $O(10^{-16})$ (CR) 
& $O(10^{-14})$ (CR) & $O(40)$ \\
n=4 & $O(10^{-8})$ (CR) & $O(10^{-8})$ (CR) & $O(10^{-6})$ (CR)  & $O(10^{-7})^*$ (CR) \\
\noalign{\smallskip}\hline\noalign{\smallskip}
\end{tabular}
\caption{Summary of typical strengths of the available constrains on the SME at different $n$ orders for rotational invariant, neutrino flavour independent LIV operators. GRB=gamma rays burst, CR=cosmic rays.
$^a$ From neutrino oscillations we have constraints on the difference of LIV coefficients of different flavors up to $O(10^{-28})$ on dim 4, $O(10^{-8})$ and expected up to $O(10^{-14})$  on dim 5 (ICE3), expected up to $O(10^{-4})$ on dim 6 op.
$^*$ Expected constraint from future experiments.  }\label{table:sum} 
\end{center}

\end{table}
 Of course at first sight this might seem a quite satisfactory state of the art, so much so that one might ask if we haven't tests Lorentz violations enough and should now move one towards new phenomenology. As usual, the answer is not a sharp one. Let us further elaborate on this point.
 
\subsection{Uncertainties on $n=4$ constraints}
Let's first stick to tests of violation of Lorentz invariance in the SME. Here, as we discussed at length in section~\ref{open-issue}, the main open issue is provided by the lasting uncertainty about the UHECR composition and heck the actual observation of the GZK cutoff. In this respect the following comment is in order. The observational picture is yes confused but not hopeless. The issue will be probably settled in a few years and some experiments like TA seem still to provide a more conservative picture than AUGER. As a matter of fact, it still seems more probable that in a few years, the constraints we presented in this review based on the GZK  cutoff will be confirmed or strengthened rather than disproved.

It would be however unfair to play down the present uncertainties in UHECR physics and place this constraints at the same level of robustness e.g. of those cast at order $n=3$ by using the synchrotron radiation from the Crab nebula. In this sense any UHECR-independent constraint (by which we mean limiting the relevant LIV parameters to be less than one, being this their natural value within the SME) on the dimension 6 operators of the SME would be more than welcomed. As we have seen in section~\ref{sec:horava}, this is possible within specific QG models, but so far only when the Lorentz breaking scale is theoretically already set to be smaller than $M_{\rm PL}$.

\subsection{The naturalness of Lorentz violations}

Another open issue is of course the naturalness problem. Lacking so far evidence for new physics at intermediate scales from the Higgs till the Planck scale one might wonder if the only hope for a working Lorentz breaking theory is gravitational confinement (see section~\ref{gravconf}) with its consequent issue of large LIV in the gravitational sector. It is too early to say, but of course investigations on specific models like the one presented in section \ref{sec:horava} seems to be strongly suggestive towards this direction. It is probably too early to say, but it is clear that the naturalness problem is probably the most pressing theoretical challenge LIV models are facing today and almost a selection tool for candidate theories which admit a low energy EFT description. Furthermore, we stress again that any emergent gravity scenario will have to be predictive in this sense as relativistic behaviour of its fundamental constituents does not seem enough, per se, to guarantee an exact relativistic emergent system \cite{Fagnocchi:2010sn}.

\subsection{Towards an authentic QG phenomenology?}

Perhaps a final comment is due to the present state of the art of the field. Lorentz breaking phenomenology has been a remarkable success, a community effort which has built (in a bit over a decade) a wealth of knowledge, methods and constraints that are now at our disposal for efficiently testing candidate theories of QG once their low energy limit is known. However, this state of affair is not and cannot be the end of the story as EFT phenomenology is not yet a true quantum gravity phenomenology. Examples like the one discussed in section~\ref{sec:horava}, should have convinced the reader that the implications of the phenomenological constraints that can be casted in full fledged QG models are more far reaching than those derived in the purely phenomenological EFT approach to LIV. We should further pursue this line of research and as ask more forcibly now to our QG models for testable low energy predictions. 

There are probably many more phenomenological consequences of QG beyond and apart LIV and maybe some of them will go beyond the realm of local EFT. For example, causal sets models seems to entail non-locality as a counterweight to the requirement of a Lorentz invariant discretisation of spacetime \cite{Dowker:2011zz}, as such the local nature of the low energy world should be in these models emergent and not exact. Can we test such non-localities? in which contexts they could be probed? Let us note again, as a side remark, that a link between some form of non-locality and extension of special relativity in situations which seem to violate the homogeneity axiom of von Ignatovski derivation have similarly been suggested  \cite{Carmona:2011wc} and probably deserve further investigations.

In conclusion, there is probably a long way to go and the challenges ahead are not just for the QG phenomenology community but also for those working on QG models. The very existence of the above described constraints on Planck scale new physics requires a new effort from these communities to search more actively for new windows for testing quantum gravity. The best has yet to come.

\ack
I wish to thank Luca Maccione, David Mattingly, Oriol Pujolas, Serguey Sibiryakov, Sebastiano Sonego and Thomas Sotiriou for useful insights, discussions and feedback on the manuscript preparation. I also thank Federico Urban for updating me on the recent evidence from the Telescope Array observations. Finally, I wish to thank Michael Romalis and Floyd Stecker for useful feedbacks on the first arXiv version of the manuscript.
\vspace{1cm}

\bibliographystyle{hunsrt}
\bibliography{references-LC}
\end{document}